\documentclass[journal]{IEEEtran}
\usepackage{cite}
\usepackage{amsmath}
\usepackage{subfigure}
\usepackage[pdftex]{graphicx}
\usepackage{amsfonts}

\usepackage{graphicx}
\usepackage{color}
\usepackage{placeins}
\usepackage{float}
\usepackage{tabularx,colortbl}
\usepackage{pbox}

\usepackage{mdwmath}
\usepackage{mdwtab}

\begin{document}

\title{Digitized metamaterial absorber-based compressive reflector antenna for high sensing capacity imaging}

\author{Ali~Molaei, Juan~Heredia-Juesas, Galia~Ghazi, James~Vlahakis and~Jose~Martinez-Lorenzo
\thanks{A. Molaei and G. Ghazi are with the Department
of Electrical and Computer Engineering, Northeastern University, Boston,
MA, 02115 USA.}
\thanks{J. Vlahakis is with the Department of Mechanical Engineering, Tufts University, Medford, MA, USA.}
\thanks{J. Heredia, and J. Martinez are with the Department
of Electrical and Computer Engineering and Department of Mechanical and Industrial Engineering, Northeastern University, Boston,
MA, 02115 USA e-mail: j.martinez-lorenzo@neu.edu.}}
\maketitle


\begin{abstract}

Conventional multistatic radar systems using microwave and millimeter-wave (mm-wave) frequencies seek to reconstruct the target in the imaging domain, employing many transmitting and receiving antenna elements. These systems are suboptimal, in that they do not take into consideration the large mutual information existing between the measurements. This work reports a new mm-wave radar system for high sensing capacity applications. The system is composed of a Compressive Reflector Antenna (CRA), whose surface is specially tailored by digitized Metamaterial Absorbers (MMAs). The MMA elements are designed to have a highly frequency-dispersive response in the operating band of the radar. This enables the CRA to create highly uncorrelated spatial and spectral codes in the imaging region. A semi-analytic method based on Drude-Lorentz model is used to approximate the reflection response of the MMAs. The performance of the developed radar system is evaluated in active mm-wave sensing systems by imaging PEC scatterers and an extended human-size model in the near-field of the radar. A computational method based on physical optics is established for solving the numerical examples. For reconstructing the image using compressive sensing techniques, a norm-1 regularized iterative algorithm based on the Alternating Direction Method of Multipliers (ADMM) and a Nesterov-based algorithm (NESTA) were applied.

\end{abstract}

\begin{IEEEkeywords}
Compressive reflector antenna, millimeter-wave imaging, metamaterial absorber, coded aperture.
\end{IEEEkeywords}

%
\IEEEpeerreviewmaketitle

\section{Introduction}

\IEEEPARstart{M}{illimeter-wave} imaging is one of the few sensing techniques that can penetrate through walls, dielectric materials, and clothing, making it a good candidate for applications such as security screening, surveillance, autonomous driving, non-destructive testing, and elsewhere \cite{Sheen2001,Martinez-Lorenzo2012,ahmed2011novel,lyons2013reflect,Martinez-Lorenzo2012a,jam2016horizontally,kharkovsky2007microwave,ferris1998microwave}. Electromagnetic waves are non-ionizing in the mm-wave frequency band, which makes the sensing harmless for radiation-sensitive targets. Unfortunately, the current sensing techniques, such as Synthetic Aperture Radar (SAR) \cite{dehmollaian2008refocusing, dehmollaian2009through}, require a mechanical raster motion to sweep the $Tx/Rx$ antennas along an effective aperture. The SAR method needs an extensive amount of time to scan the imaging domain, making real-time imaging impractical. The development of a fast, fully electronic system does not require a mechanical scanning approach; however, implementing such system would be highly expensive due to the large number of $Tx/Rx$ radar modules required \cite{Martinez-Lorenzo2012}. 

Recently, several fully electronic-based imaging systems have been proposed, aiming to perform the sensing faster (in real-time) and more efficiently. A frequency diverse one-dimensional radar, which is based on an active metamaterial aperture, is reported in \cite{sleasman2016design}. The imaging system operates in the microwave frequencies ($17.5$-$21.1$ \textit{GHz}). The metamaterial elements are dynamically controlled by voltage-controlled active diodes, producing a series of distinct radiation patterns that illuminate the imaging region. Due to the increased size and bandwidth of the active aperture, the obtainable information is increased and real-time imaging is achieved. 
As another example, in \cite{ahmed2009multistatic} a multistatic mm-wave imaging system with planar 2D-arrays is presented. A digital beamforming technique is used to reconstruct the images. Compared to the monostatic setup, the proposed array configurations (grid array and plus array) provide a considerably better dynamic range.

In this work, a specially coded CRA \cite{Martinez-Lorenzo2015,MartinezLorenzo2015,molaei2018interferometric,juesas2015consensus,molaei2016interferometric,molaei2016active,molaei20172,molaei2017high,heredia2017norm,molaei2017compressive}, which is based on digitized MMAs, is proposed to be used for mm-wave sensing systems. 
The CRA has the ability to illuminate the imaging region with spatial and spectral coded patterns. The codes that will enable high bandwidth communications will be dynamically assigned while maximizing the overall performance of the network. The proposed mm-wave sensing system can be regarded as a coded aperture that modifies the effective Singular Value (SV) spectrum seen by the transmitting and receiving arrays. Contrary to the systems described above, the proposed CRA-based system can be reconfigured to operate in the far-field, thus enabling standoff detection of security threats.

The paper is organized as follows: Section 2 describes the framework of the system and its underlying parameters. This is done through the following steps: (a) defining a linear equation that expresses the physics of the system; (b) describing the sensing capacity as a parameter to quantify the amount of information received by a sensing system.  Section 3 is dedicated to the design and characterization of the digitized binary MMA: first, the polarization-independent meander-line MMA as a unit-cell element is introduced; next, the digitized MMAs built from the unit-cell meander-line elements, along with their frequency dispersive response, are presented; finally, the digitized MMAs are characterized based on a semi-analytical formulation derived from a multi-resonance Drude-Lorentz model. In Section 4, the performance of the system is validated through a numerical example for sensing metallic scatterers. The results of this testing are reported to demonstrate the performance of the system in terms of sensing capacity, radiation pattern, beam focusing, and image reconstruction. Section 5 describes a CRA-based array, composed of eight compressive reflectors, to image an extended human-size region. The surface of each CRA is tailored by digitized MMAs to enhance the sensing capacity of the array imaging system. Finally, the paper is concluded in Section 6 by discussing some extensions to the presented imaging systems.

\section{Metamaterial Absorber-Based Compressive Reflector Antenna}

\begin{figure}[htp]
	\centering
	\includegraphics[width=0.47\textwidth,trim={0cm 1.3cm 1.3cm 2.5cm},clip=true]{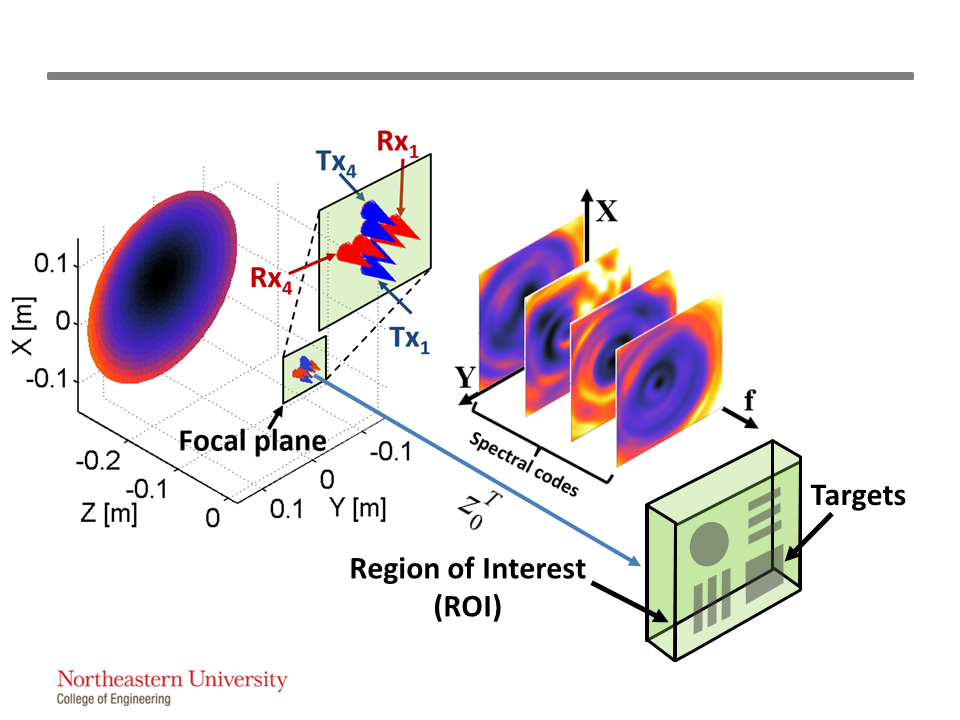}
	\caption{Illustration of the proposed imaging setup: feeding array located on the focal plane induces currents on the surface of the CRA, which ultimately creates the spectral codes in the imaging region. As an example of these spectral codes, magnitude of electric field created by the CRA when illuminated by $Tx_2$ is shown as a function of frequency.}
	\label{Geo}
\end{figure}

The proposed MMA-based CRA system, as illustrated in Fig. \ref{Geo}, is composed of a custom designed reflector antenna with a 2D array of conical horns as feeding elements. The conical horns are configured as a \textit{plus shape} along the focal plane of the reflector with the following arrangement: $N_{Tx}$ transmitting and $N_{Rx}$ receiving elements are positioned equi-distantly along the $x$-axis and $y$-axis, respectively; the antenna elements feed the reflector by a linear polarization in the $y$-axis. 

\begin{figure}[htp]
	\centering
	\includegraphics[width=0.3\textwidth,trim={3.7cm 0.58cm 5cm 2.5cm},clip=true]{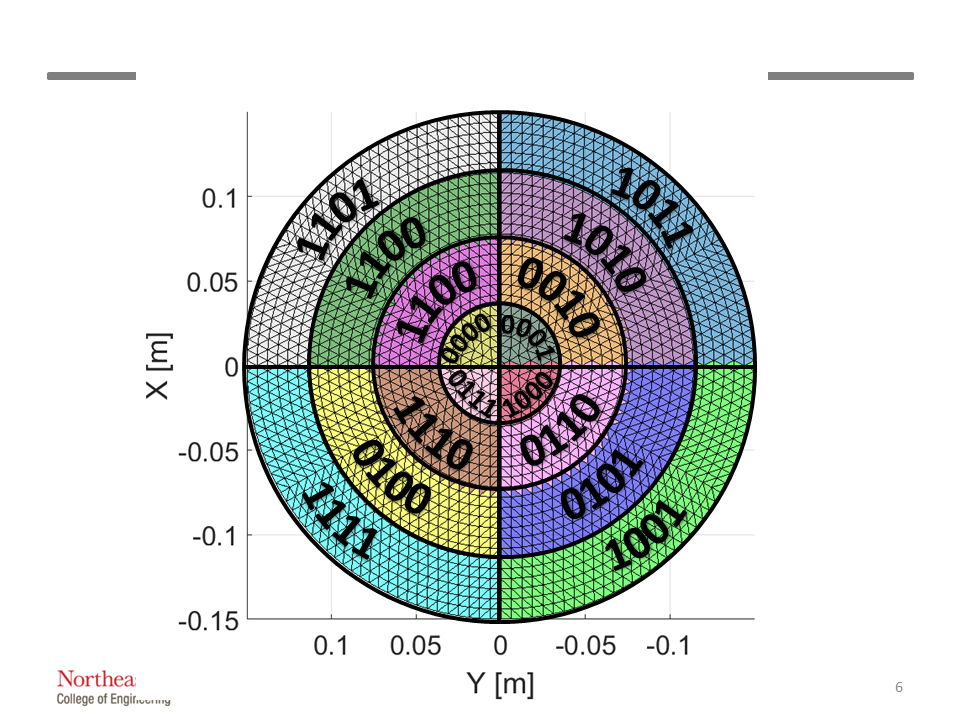}
	\caption{Front view of the proposed CRA: surface of the reflector is divided into 16 domains and each domain is coated with a unique 4-bit MMA.}
	\label{DigitizedMMA}
\end{figure}

The MMA-based CRA consists of a Traditional Reflector Antenna (TRA), whose surface is covered with digitized MMAs. In a general case, an $n$-bit binary code $c_i=a_1a_2...a_n$, $i \in \{1 ,...,2^n \}$, is associated to each of the MMA designs. Each digit of the binary code $a_j \in \{0,1 \}$, where $j \in \{ 1,2,...,n \}$, is associated with a resonance frequency $f_j$ within the operating frequency band. Depending on the binary value `0' or `1' associated to each digit $a_j$, the MMA elements are designed to {\it reflect} or {\it absorb} the incident field at the resonance frequency $f_j$. Figure \ref{DigitizedMMA}, for example, shows the surface of a TRA divided into 16 domains and each domain is coated with a unique 4-bit MMA. 


The rest of this section is dedicated to describe the framework of the system and its underlying parameters. In subsection \ref{SensingMatrix}, a linear equation that expresses the physics of the system is defined. Then, in subsection \ref{SensingCapacity}, the sensing capacity as a parameter to quantify the amount of information received by the sensing system is described.

\subsection{Sensing Matrix} \label{SensingMatrix}

Under the configuration mentioned previously, each receiver collects the signal from each transmitter for $N_f$ different frequencies, for a total number of $N_m=N_{Tx}\cdot N_{Rx} \cdot N_f$ measurements. The image reconstruction domain is made up of $N_p$ pixels in the Region Of Interest (ROI) located $z_0^T$ meters away from the focal plane of the reflector.
Under this configuration, the vector of complex field measured data $\textbf{g} \in {\mathbb{C}}^{N_m}$ is related to the unknown complex vector of reflectivity $\textbf{u} \in {\mathbb{C}}^{N_p}$, defined in each pixel, through the sensing matrix $\mathbf{h}_i = \mathbf{E}_{T_x^i} \circ \mathbf{E}_{R_x^i}$. The vector $\mathbf{h}_i$ is the $i$-th row of $\textbf{H}$, $\mathbf{E}_{T_x^i}$ and $\mathbf{E}_{R_x^i}$ are, respectively, the fields from the \textit{Tx} and \textit{Rx} on the $i$-th measurement, and the symbol $\circ$ is the Hadamard product.  

This relationship can be expressed in a matrix form as follows:
\begin{equation}
\bf{g=Hu+w},
\label{sensing_eq}
\end{equation}
where $\textbf{w}\in\mathbb{C}^{N_m}$ represents the noise collected at each measurement.

\subsection{Sensing Capacity} \label {SensingCapacity}

The maximum information rate has been studied by the field of information theory. It was pioneered by Shannon, almost seventy years ago \cite{shannon1948mathematical}. A similar concept of \textit{\textbf{Sensing Capacity}} with the aim to quantify the amount of information received by a sensing system has been studied in \cite{Martinez-Lorenzo2015}, and it is defined as follows:

\begin{equation}\label{eq:capacity}
C_{\epsilon}^{M} \approx \log_2 \left ( \prod_{m=1}^{M}\frac{\sigma_{m}^{\mathbf{H}}\cdot e_m}{\epsilon} \right )=\sum_{m=1}^M\log_2\left(\frac{\sigma_m^{\bf{H}}\cdot e_m}{\epsilon}\right)\;[bits]
\end{equation}
where $\sigma_{m}^{\mathbf{H}}$ is the $m$-th Singular Value (SV) of the sensing matrix ${\mathbf{H}}$, $e_m$ is power of the signal received in the $m$-th orthogonal channel of ${\mathbf{H}}$, and $M$ is the number of SVs that are above the uncertainty level $\epsilon$.

According to Eq. (\eqref{eq:capacity}), the sensing capacity of a system operating under an uncertainty level $\epsilon$ and a total transmitted power $e$ can be increased in two ways: 1) by reducing the dispersion of the SVs, their spectrum becomes flatter, so more SVs can be measured above the noise level; and 2) by increasing the amount of received power $e_r=\sum_{m=1}^M{e_m}=e\cdot\eta_r$, where $\eta_r \leq 1 $ refers to the fraction of the transmitted power that is measured by the receiving array.

\section{Design and Characterization of Digitized Binary Metamaterial Absorbers} \label{Sec_MMA}

The goal in this section is to present the procedure for the design of the digitized MMA. Subsection \ref{sub1} begins by developing a polarization-independent unit-cell, followed by a 4-bit MMA design. 
Subsection \ref{sub2} focuses on the design of 16 different MMA configurations. These MMAs are later used for building a diverse set of frequency dispersive codes, by being used as a coating on the surface of the reflector.

\subsection{Quasi Polarization-Independent Meander-Line Metamaterial with Near-Unity Absorption} \label{sub1}

\begin{figure}[ht!]
    \centering
        \subfigure[]{\label{MMA-Fab-a}
        \includegraphics[width=0.35\textwidth,trim={0cm 4cm 6.3cm 4.55cm},clip=true]{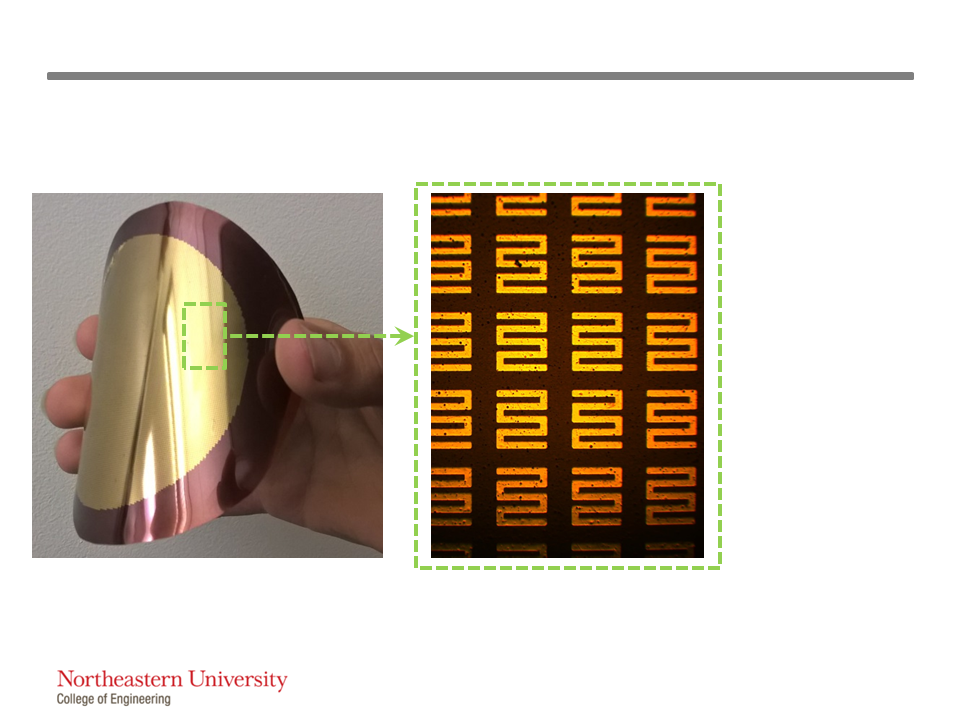}}
				~
				\subfigure[]{\label{MMA-Fab-b}
        \includegraphics[width=0.35\textwidth,trim={0cm 0cm 0cm 0cm},clip=true]{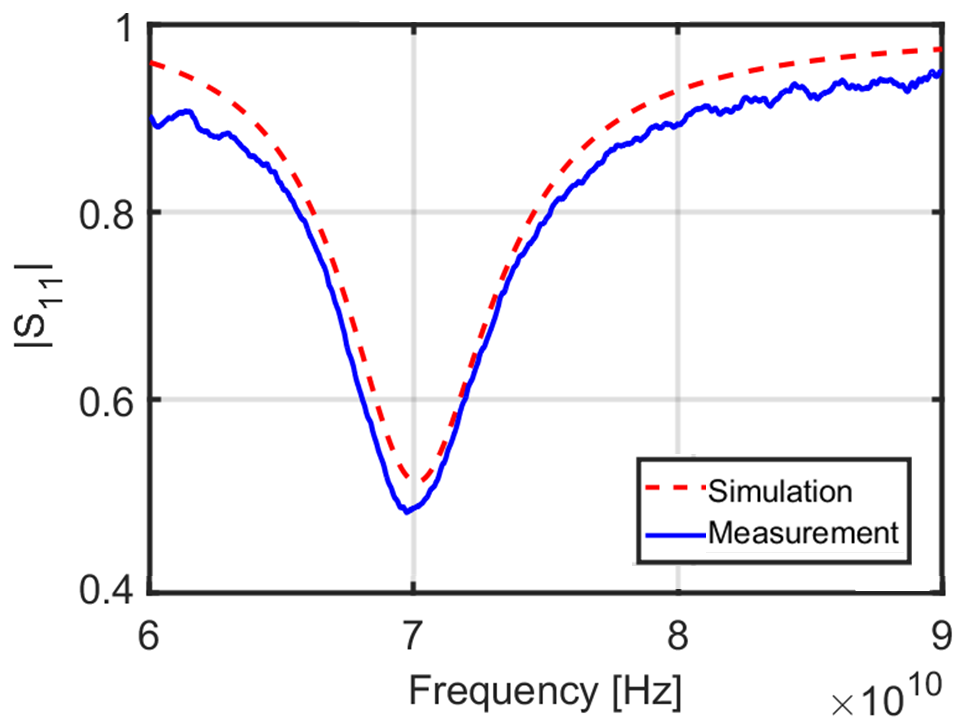}}

    \caption{ \label{MMA-Fab} (a) Fabricated meander-line MMA on a polyimide substrate: For fabricating the MMA array, gold sputtering followed by a lift-off process is used to pattern the top layer. For the bottom layer, only a gold sputtering process is performed. Based on the simulation $20/200nm$ of \textit{Cr/Au} was targeted, but based on the post-deposition measurement, the actual deposition is estimated to be slightly less ($18/185nm$ \textit{Cr/Au}). (b) Simulation and measurement comparison of the magnitude of the $S_{11}$. }
\end{figure}

In the work presented here, a previously developed MMA called meander-line \cite{watts2015metamaterials} is utilized and modified to build the unit-cell. Fabrication of the meander-line MMA presented in \cite{watts2015metamaterials} is performed using gold sputtering followed by a lift-off process [Fig. \ref{MMA-Fab-a}]. Three different meander-line MMAs having resonance frequencies in the range of 70-77 \textit{GHz} are designed and fabricated. As an example, simulation and measurement comparison of the magnitude of the $S_{11}$, corresponding to the meander-line MMA resonating at 70 \textit{Ghz}, is shown in Fig. \ref{MMA-Fab-b}. For the purpose of this work, the meander-line unit-cell is modified to design a new quasi polarization-independent MMA.

\begin{figure}[ht!]
    \centering
        \subfigure[]{\label{SingleUnit-a}
        \includegraphics[width=0.25\textwidth,trim={4cm 1.5cm 4cm 2.4cm},clip=true]{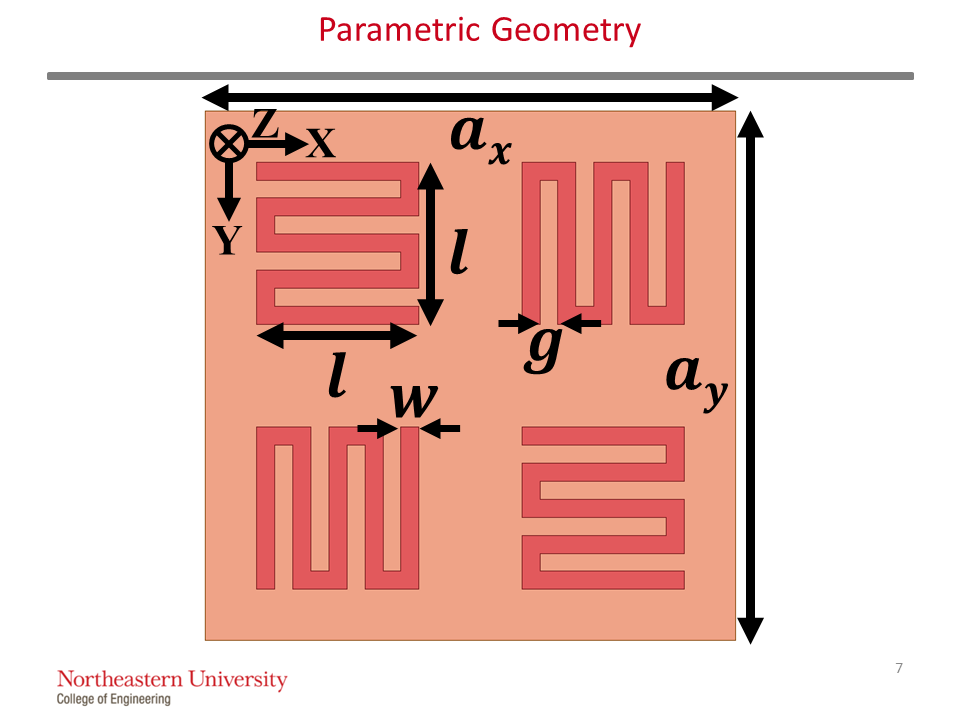}}
				~
				\subfigure[]{\label{SingleUnit-b}
        \includegraphics[width=0.35\textwidth,trim={0.2cm 0.1cm 1.2cm 4cm},clip=true]{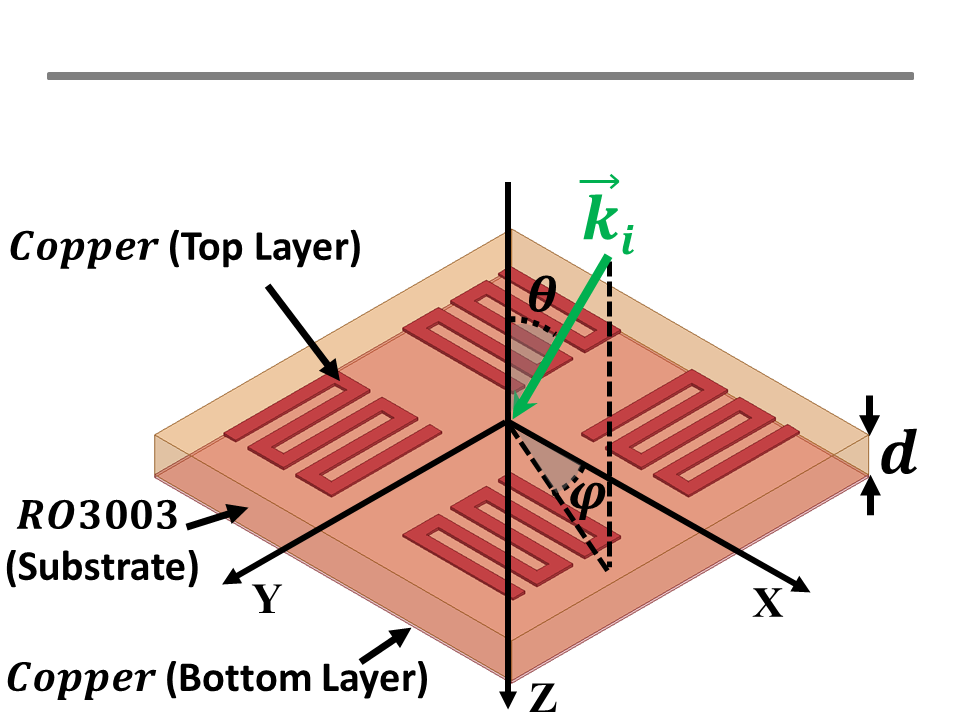}}

    \caption{ \label{SingleUnit} (a) Top view, and (b) 3-D view of the polarization-independent meander-line MMA unit cell. The dimensions in $[\mu m]$ are: $a_x=a_y=1300,$ $l=398$, $w=43$, $g=43$. }
\end{figure}

The MMA unit-cell is simulated and designed using HFSS, a commercial software based on Finite Element Analysis (FEA). Master/slave periodic boundary conditions are defined for the $xz$- and $yz$-planes surrounding the unit-cell. Figure \ref{SingleUnit} shows the unit-cell structure of the presented polarization-independent meander-line MMA. A general incident plane wave with an arbitrary elevation ($\theta$) and azimuth ($\phi$) angles is defined to excite the unit-cell [Fig. \ref{SingleUnit-b}].

The geometric parameters of the MMA, which is designed to resonate at $71$ \textit{GHz}, are given by $a_x=a_y=1300$, $l=398$, $w=43$ and $g=43$, all in $\mu m$. Rogers RO3003 with a thickness $d$ of $127$ $\mu m$ is used as the substrate, whose permittivity and dielectric loss tangent are $3$ and $0.0013$, respectively. The meander-line patterns on the top layer and the metal coating on the bottom layer are composed of copper with a thickness of $17$ $\mu m$.

The unit-cell benefits from two unique features: ($i$) it is almost insensitive to the polarization of the incident plane wave, i.e. for a given elevation incident angle ($\theta$ in Fig. \ref{SingleUnit-b}), the reflection coefficient is not significantly altered for different azimuth incident angles ($\phi$ in Fig. \ref{SingleUnit-b}), ($ii$) it has a near-unity absorption rate for wave excitations with normal incident angles, i.e. for the elevation incident angle of $\theta = 0$, the absorption rate value, which is defined as $A = 1- |S_{11}|^2 - |S_{21}|^2$, is near unity. It is noteworthy to mention that we only need the $S_{11}$ term to calculate the absorption rate, since the bottom layer of the MMA is covered by copper [Fig. \ref{SingleUnit-b}] and there would be no transmission through the MMA array, therefore $|S_{21}|=0$.

\begin{figure}[ht!]
    \centering
        \subfigure[]{\label{TE}
        \includegraphics[width=0.4\textwidth,trim={2cm 1.5cm 3cm 3cm},clip=true]{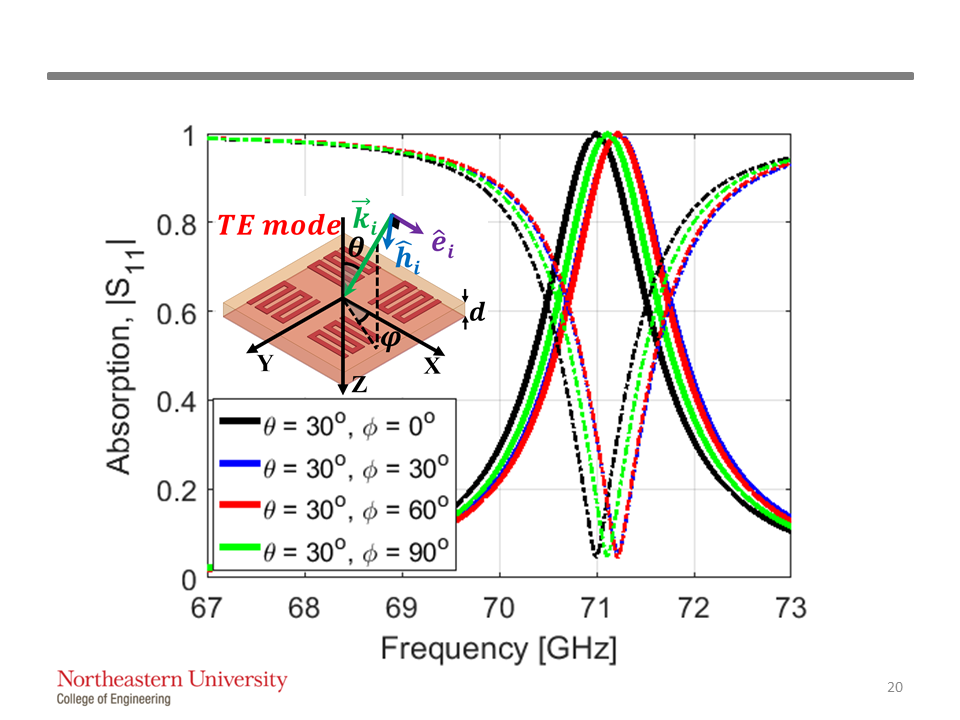}}
				
				\subfigure[]{\label{TM}
        \includegraphics[width=0.4\textwidth,trim={2cm 1.5cm 3cm 3cm},clip=true]{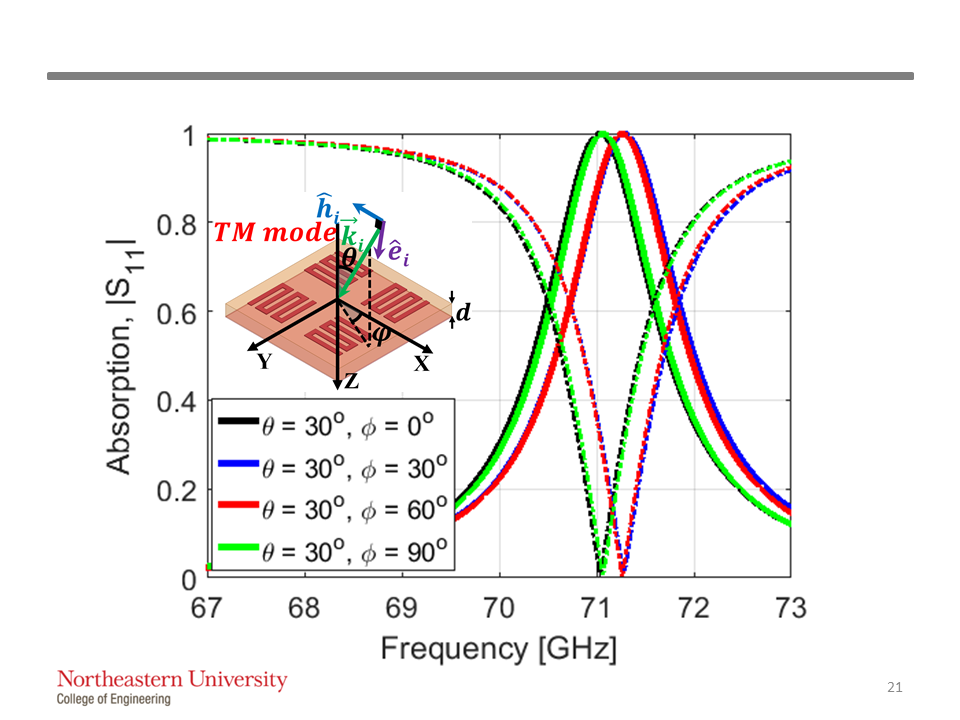}}

    \caption{ \label{Pol_ind} Absorption rate (solid lines) and $|S_{11}|$ (dashed lines) of the quasi polarization-independent MMA at various
incident angles for the (a) TE, and (b) TM incidence.}
\end{figure}

A set of simulations was carried out to justify the aforementioned features for the proposed unit-cell. Figs. \ref{TE} and \ref{TM} illustrate the absorption rate and $|S_{11}|$ plots for TE and TM excitations, respectively. The plots are for a fixed value of elevation ($\theta=30^{\circ}$) and different azimuth ($\phi=0^{\circ}, 30^{\circ}, 60^{\circ}$ and $90^{\circ}$ ) incident angles. It is apparent from the absorption rate plots that, for different azimuth incident angles, there is a small shift in the frequency response (less than $0.3\%$), and the peak at resonance frequencies is close to unity (larger than $0.99$). The absorption bandwidth, defined as the frequency range in which absorption is above half its maximum value, is $1.64$ \textit{GHz}.

\begin{figure}[ht!]
	\centering
	\includegraphics[width=0.45\textwidth,trim={1cm 2cm 2cm 3cm},clip=true]{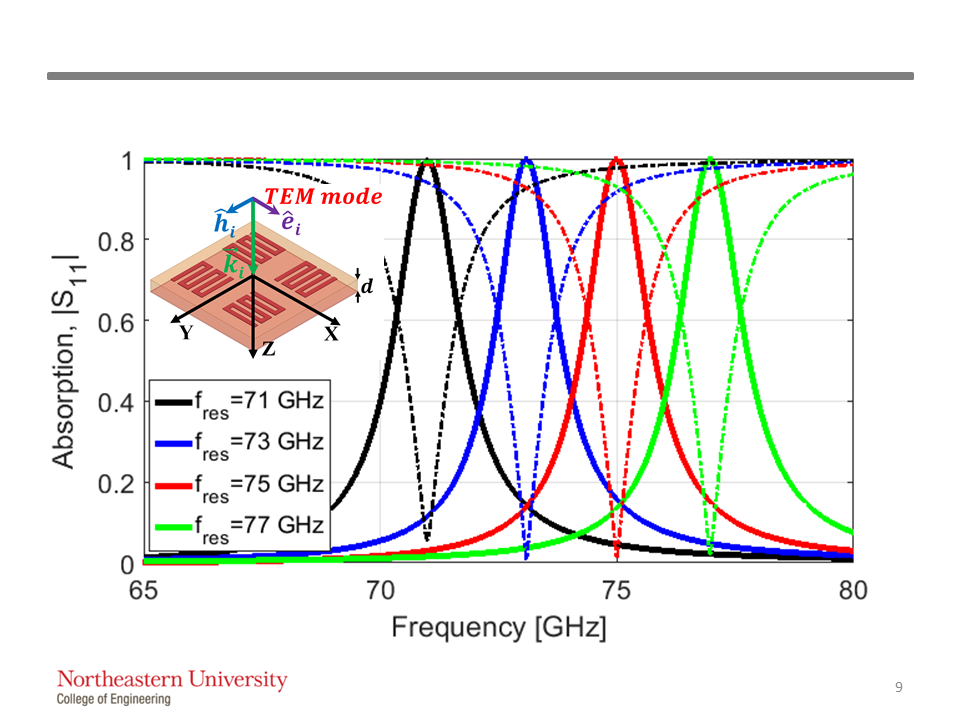}
	\caption{ Absorption rate (solid lines) and $|S_{11}|$ (dashed lines) of the MMAs with different scaling factors, for the TEM excitation: (black) $S=1.00$, (blue) $S=0.97$, (red) $S=0.95$, and (green) $S=0.93$.}
	\label{FourRes}
\end{figure}

A scaling factor to the original MMA design can be applied, in order to obtain different designs resonating at different frequencies. In this regard, the original design, whose center resonating frequency is $71$ \textit{GHz}, is assumed to have the base dimensions with scaling factor $S$ equal to 1. Three additional designs with center resonance frequencies of $73$ \textit{GHz}, $75$ \textit{GHz}, and $77$ \textit{GHz} are configured by adjusting the scaling factor to be $S=0.97$, $S=0.95$, and $S=0.93$, respectively. For the new designs, the periodicities of the unit-cell ($a_x$ and $a_y$) are kept the same and the scaling is applied only on the geometrical parameters of the meander-line pattern ($l$, $w$ and $g$). Figure \ref{FourRes} depicts the absorption rate and $|S_{11}|$ for the four MMA designs, which are excited by a \textit{TEM}-mode plane wave.

\subsection{Design of a 4-bit Metamaterial Absorber} \label{sub2}

\begin{figure}[ht!]
    \centering
        \subfigure[]{\label{Digit-a}
        \includegraphics[width=0.26\textwidth,trim={1cm 1.4cm 1cm 2.2cm},clip=true]{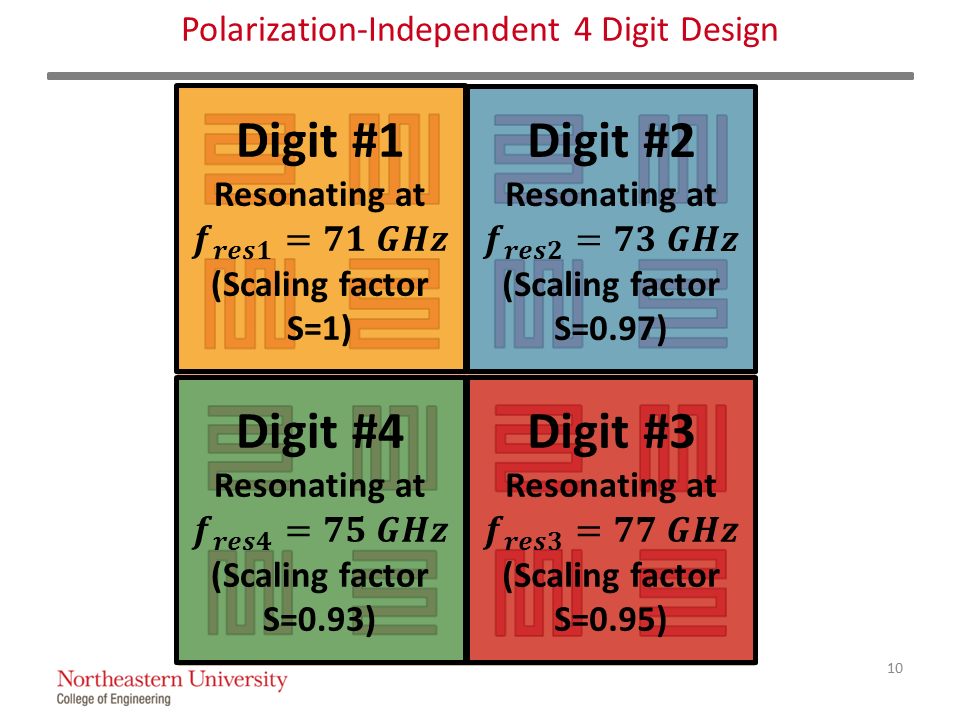}}

				\subfigure[]{\label{Digit-b}	
        \includegraphics[width=0.37\textwidth,trim={1.5cm 4.7cm 2cm 4cm},clip=true]{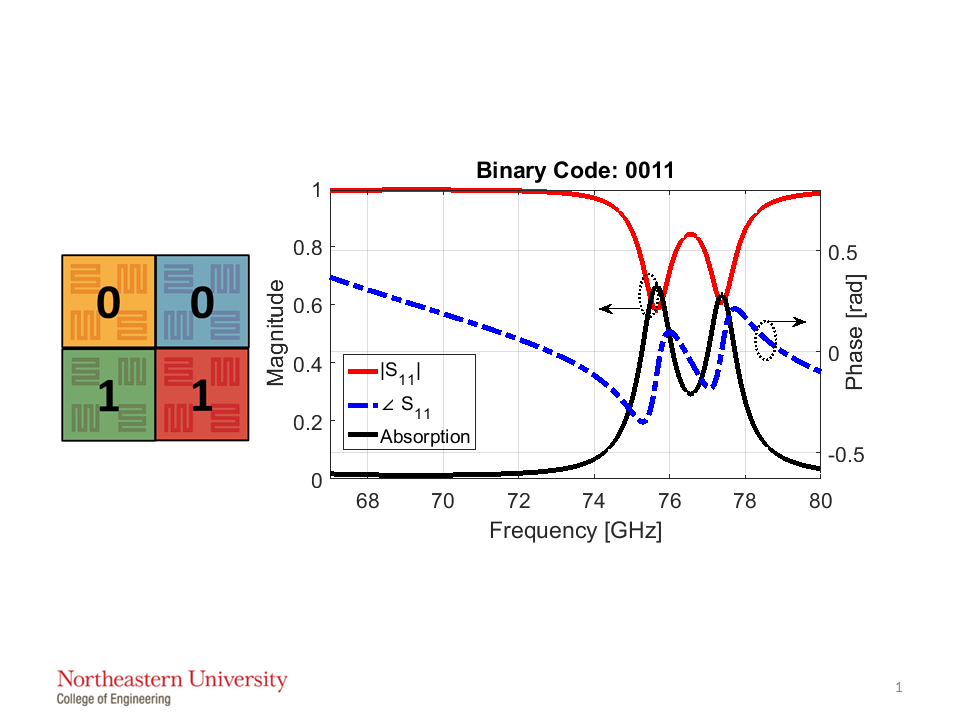}} \vspace{-0.38cm}

        \subfigure[]{\label{Digit-c}
        \includegraphics[width=0.37\textwidth,trim={1.5cm 4.7cm 2cm 4cm},clip=true]{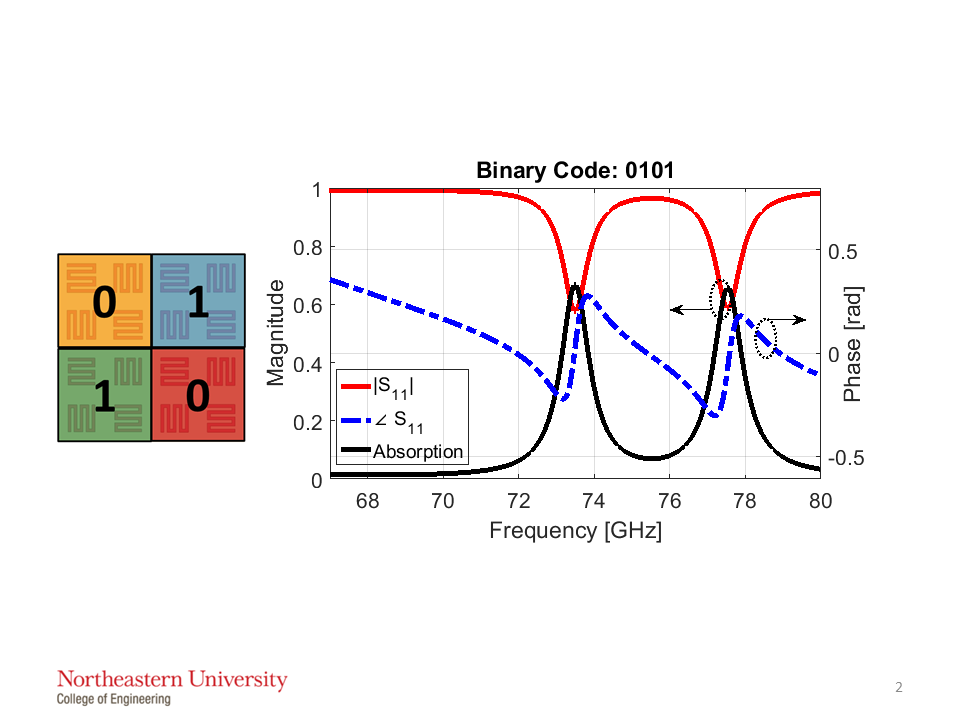}} \vspace{-0.38cm}

				\subfigure[]{\label{Digit-d}
        \includegraphics[width=0.37\textwidth,trim={1.5cm 4.7cm 2cm 4cm},clip=true]{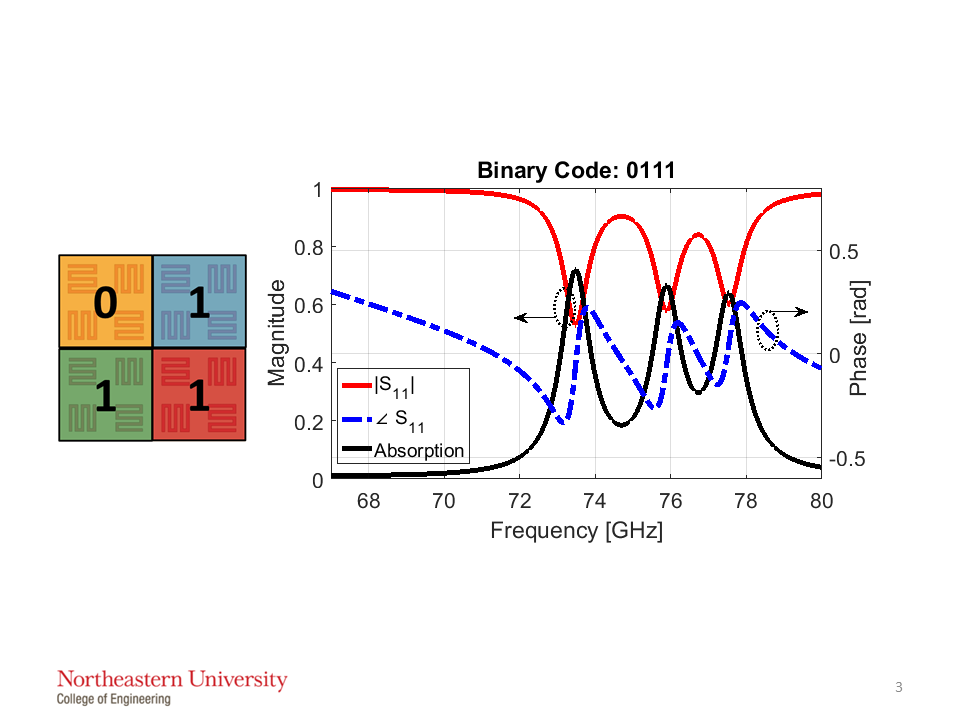}} \vspace{-0.38cm}

				\subfigure[]{\label{Digit-e}
        \includegraphics[width=0.37\textwidth,trim={1.5cm 4.7cm 2cm 4cm},clip=true]{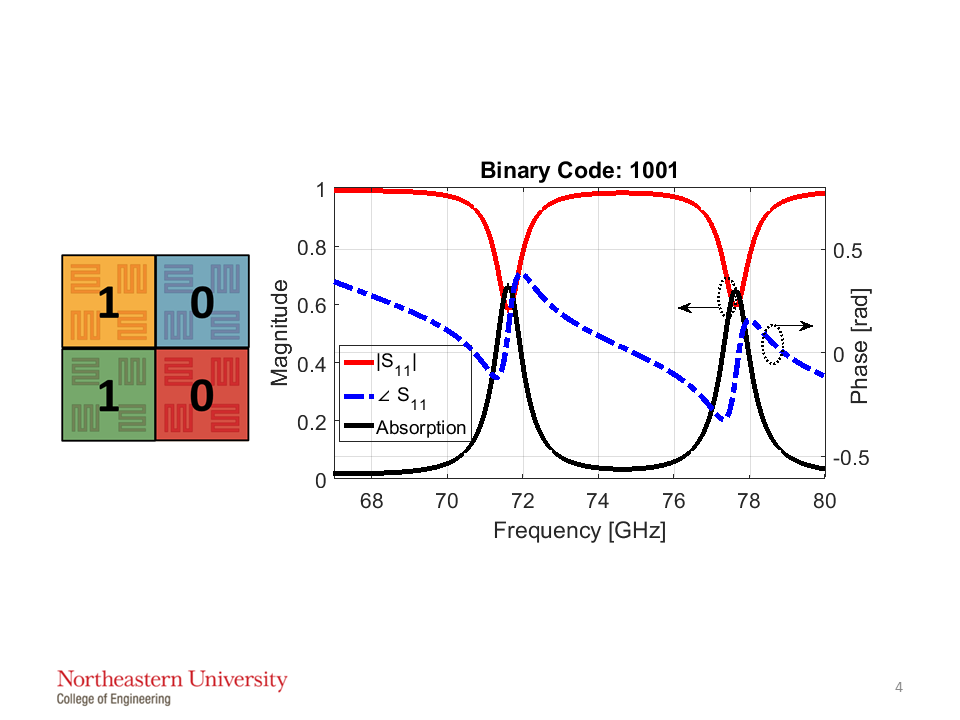}} \vspace{-0.38cm}

				\subfigure[]{\label{Digit-f}
        \includegraphics[width=0.37\textwidth,trim={1.5cm 4.7cm 2cm 4cm},clip=true]{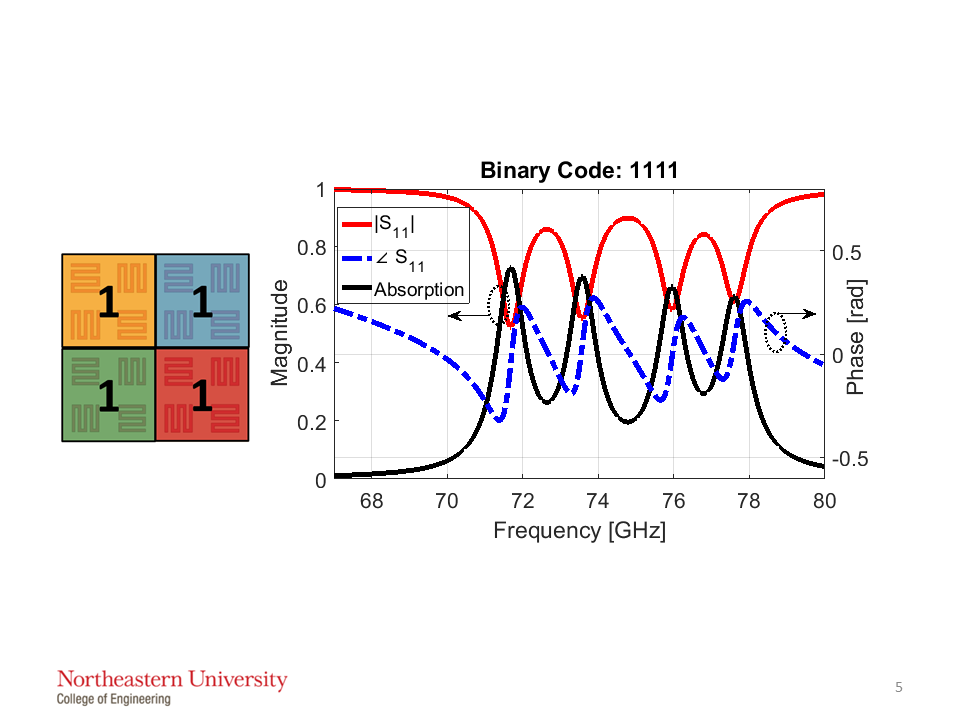}} 

    \caption{ \label{Digit} Examples of absorption rate and $S_{11}$ of the (a) 4-bit MMA unit cell with binary digits of (b) $c_3$=`0011', (c) $c_5$=`0101', (d) $c_7$=`0111', (e) $c_9$=`1001', (f) $c_15$=`1111'.}
\end{figure}

The unit-cell design presented in subsection \ref{sub1} is extended to a $2 \times 2$ configuration, as illustrated in Fig. \ref{Digit-a}, to design a digitized 4-bit MMA ($n=4$).
Each digit $a_j$ of the 4-bit MMA, is associated with one of the four basic MMA unit-cell designs. Based on the `1' or `0' values associated with each digit of the 4-bit binary code $c_i$, the MMA may or may not absorb the impinged electromagnetic power at four different resonance frequencies, respectively. The  scaling factor $S$ for the $a_1$, $a_2$, $a_3$, and $a_4$ digits corresponds to resonances at $f_1=71$ $GHz$, $f_2=73$ $GHz$, $f_3=75$ $GHz$, and $f_4=77$ $GHz$, respectively. For the `0' value corresponding to a digit, the scaling factor for that digit is selected to have an out-of-band resonance ($S=1.1$). Simulations show that the bandwidth for the single resonance MMAs is about $1.56$ \textit{GHz}, equivalent to an approximate quality factor of $46$. Having a higher quality factor, more resonances can be included in the same operating frequency range, therefore, more spectral codes can be added to the imaging system. Here, based on the achieved quality factor for the single resonance MMAs, up to four resonances were used in the $70-77 GHz$ frequency range.

Figs. \ref{Digit-b}-\ref{Digit-f} illustrate the absorption rate, magnitude, and phase of $S_{11}$ plots for some examples of the presented 4-bit MMA. A \textit{TEM}-mode plane wave excitation is applied for these examples. It is apparent from the plots that there is some frequency shift in the peak value of the absorption rate with respect to the expected resonance frequency. This is due to the mutual coupling between the adjacent meander-line elements. However, the scaling factors for each digit could be customized to compensate the unwanted mutual coupling and have the absorption rate peak at the desired frequencies.

\begin{figure}[ht!]
	\centering
	
        \subfigure[]{\label{TEM_Geo}
        \includegraphics[width=0.22\textwidth,trim={2cm 0cm 1cm 2.3cm},clip=true]{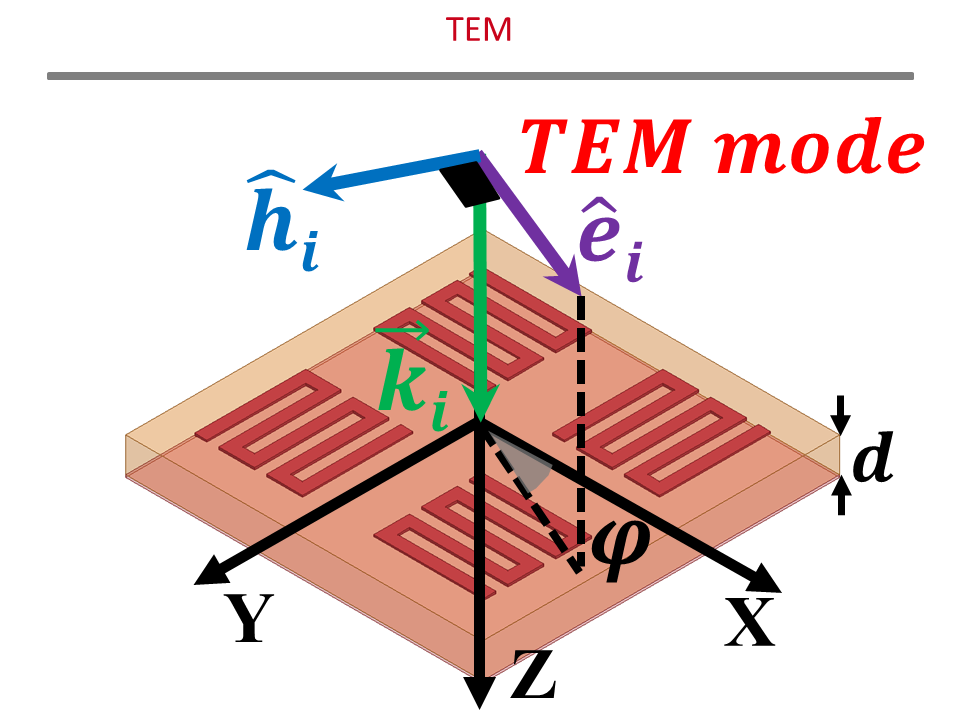}}

				\subfigure[]{\label{Phi_0}
        \includegraphics[width=0.41\textwidth,trim={2cm 1.5cm 2cm 2.3cm},clip=true]{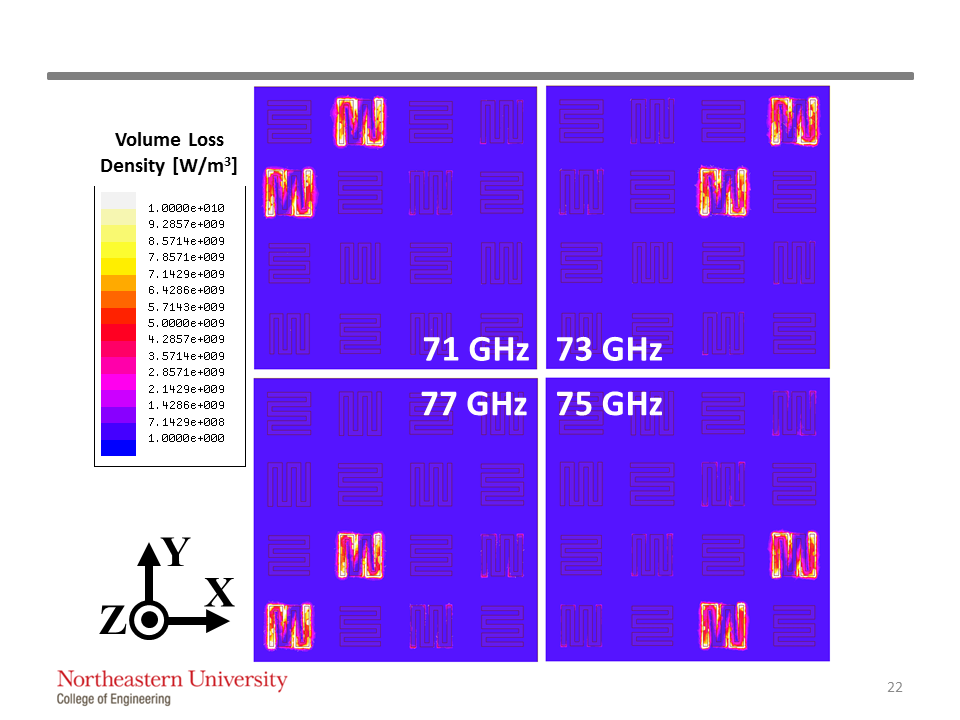}}

        \subfigure[]{\label{Phi_45}
        \includegraphics[width=0.41\textwidth,trim={2cm 1.5cm 2cm 2.3cm},clip=true]{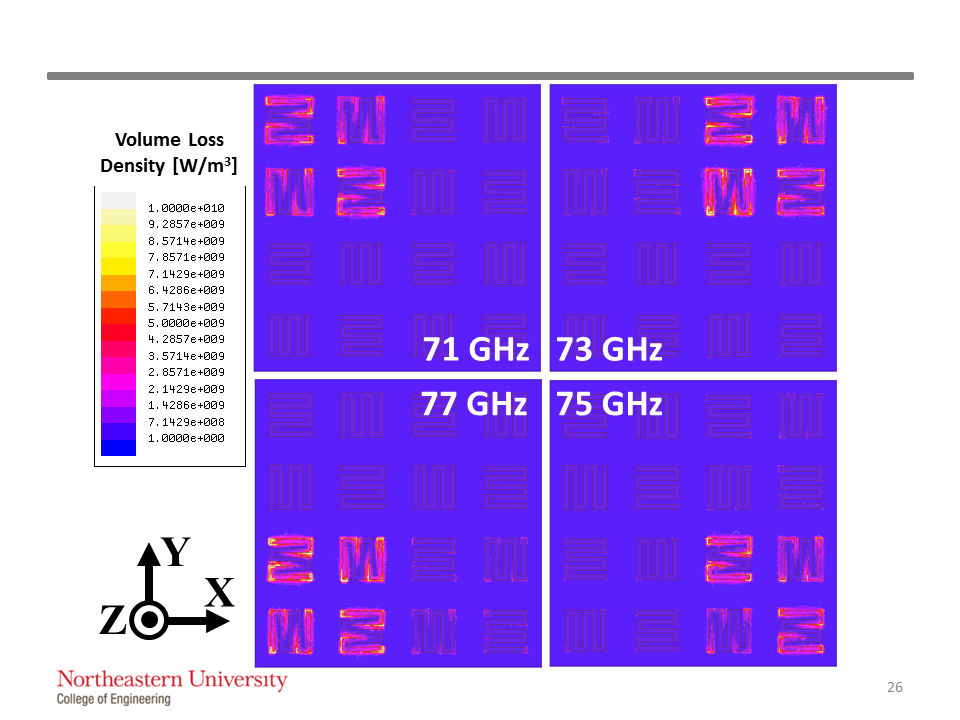}}

				\subfigure[]{\label{Phi_90}
        \includegraphics[width=0.41\textwidth,trim={2cm 1.5cm 2cm 2.3cm},clip=true]{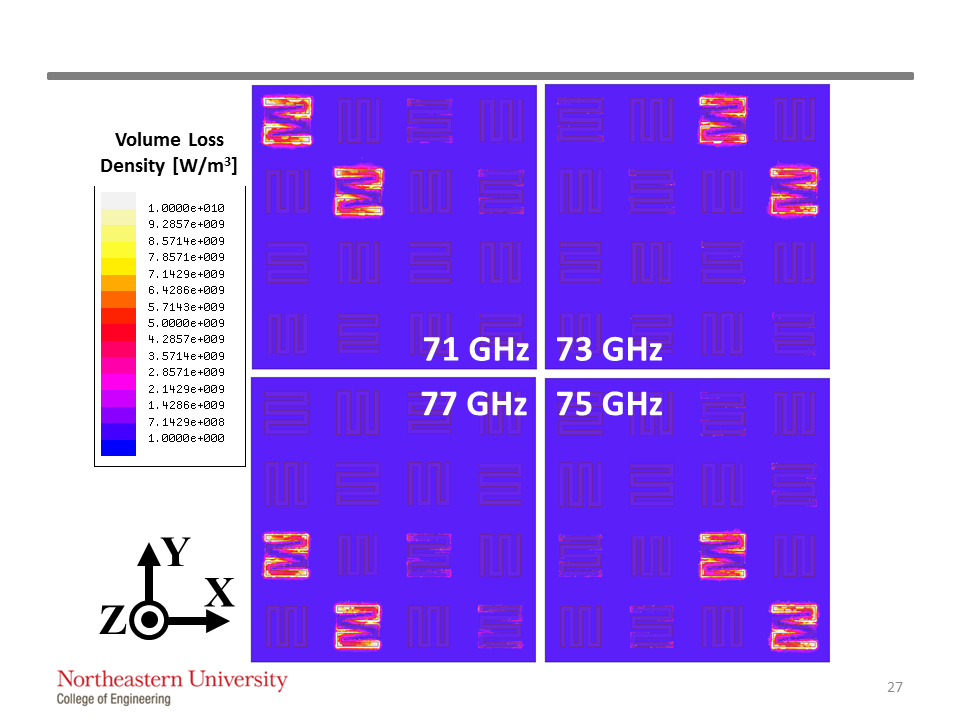}}

	\caption{ \label{Volume_loss} (a) TEM excitation of the 4-bit MMA unit cell with binary digit $c_{16}=$`1111'. Volume loss density on top layer of the MMA unit cell with the configuration shown in (a), for (b) $\phi = 0^{\circ}$, (c) $\phi = 45^{\circ}$, and (d) $\phi = 90^{\circ}$.}
\end{figure}

In order to further understand the behavior of the 4-bit MMA, volume loss density for one unit-cell of the array has also been studied. Figure \ref{Volume_loss} depicts the volume loss density on the top substrate layer of the 4-bit MMA unit cell with binary code number $c_{16}=$`1111', excited by a \textit{TEM}-mode plane wave and an azimuth angle of $\phi = 0^{\circ}$ [Fig. \ref{Phi_0}], $\phi = 45^{\circ}$ [Fig. \ref{Phi_45}], and $\phi = 90^{\circ}$ [Fig. \ref{Phi_90}]. As illustrated in Fig. \ref{Phi_0}, the E-field has a component only in the $x$-axis and, consequently, only the meander-line elements aligned in the $y$-axis have absorbed the incident power. However, in Fig. \ref{Phi_45}, the E-field has components in both the $x$-axis and $y$-axis; hence, both meander-line elements aligned in the $x$-axis and $y$-axis have absorbed the incident power. Finally, in Fig. \ref{Phi_90}, the E-field has a component only in the $y$-axis and, as a result, only the meander-line elements aligned in the $x$-axis have absorbed the incident power. Moreover, it can be seen from Fig. \ref{Volume_loss} that, as expected, at 71 \textit{GHz}, 73 \textit{GHz}, 75 \textit{GHz}, and 77 \textit{GHz}, only the meander-line elements associated with digits $a_1$, $a_2$, $a_3$, and $a_4$ have absorbed the incident power, respectively.

Keeping the periodicity ($a_x$ and $a_y$) constant might break the symmetry; however, doing so will not restrict the performance of the design for the following reasons: (a) Other geometrical parameters ($l$, $g$, and $w$) can be optimized to shift the resonance frequency; and (b) the change in the scaling factor ($S$) is very small for different designs, so it can be assumed that there is not a large discontinuity in the symmetry related to the periodicity of the unit-cells. The main reason that the periodicity has been kept constant is to be able to easily place the MMA digits side-by-side, to build a larger cell with a well-defined square shape. In this way, digitized MMA apertures could be engineered on a 2D surface without having a gap between the MMA elements.


\subsection{Characterization of the MMA, Based on Multi-Resonance Drude-Lorentz Model}

A semi-analytical model is used to characterize the electromagnetic response of the meander-line MMAs. The characterization is performed in terms of the total reflection coefficient of the meander-line MMAs. Each MMA is modeled as a bulk magneto-dielectric medium \cite{Watts2012} and characterized by a generalized multi-resonance Drude-Lorentz model \cite{Ugawa1999}. This generalized model allows any number of resonances to be accounted for with a simple summation:
\begin{subequations} \label{DrudeLorentz}
\begin{align}
\tilde{{\varepsilon}}_{r}(\omega )=\varepsilon_{\inf} + \omega^{2}_{p,e}\sum_{k=1}^{N}\frac{f_{k,e}}{\omega^{2}_{0,k,e}-\omega^{2}-i\gamma_{k,e}\omega}\\
\tilde{{\mu}}_{r}(\omega )=\mu_{\inf} + \omega^{2}_{p,m}\sum_{k=1}^{N}\frac{f_{k,m}}{\omega^{2}_{0,k,m}-\omega^{2}-i\gamma_{k,m}\omega}
\end{align}
\end{subequations}
in which $\varepsilon_{\inf}$ and $\mu_{\inf}$ are the static
permittivity and permeability at infinite frequency, respectively; $N$ is the total number of resonances; 
$\omega_{p,e}$ and $\omega_{p,m}$ are the plasma frequencies;
$\omega_{0,k,e}$ and $\omega_{0,k,m}$ are the resonant frequencies of the $k_{th}$ resonance;
$\gamma_{k,e}$ and $\gamma_{k,m}$ are the damping frequencies of the $k_{th}$ resonance; and 
$f_{k,e}$ and $f_{k,m}$ are the oscillator strengths of the $k_{th}$ resonance.

The semi-analytical model is developed from optimizing the generalized Drude-Lorentz parameters in Eq. \eqref{DrudeLorentz} of a three-layer magneto-dielectric media, with the goal of having a reflection coefficient equal to that computed using the commercial FEA method software HFSS. The total reflection coefficient from the stratified three-layer magneto-dielectric medium is given by \cite{molaei2018interferometric}: 

\begin{align} \label{Gamma2}
\begin{split}
\Gamma &= \Gamma_{12}+\frac{T_{12}\Gamma_{23}T_{21}e^{-j2\Phi ^{trans}}}{1-\Gamma_{23}\Gamma_{21}e^{-j2\Phi ^{trans}}}\\
\end{split}
\end{align}
in which $\Gamma$ is the total reflection coefficient; $\Gamma_{ij}$ and $T_{ij}$ are the reflection and transmission coefficients, respectively, linked to the interface between medium $i$ and medium $j$; and $\Phi^{trans}$ is a complex value that accounts for the phase delay and amplitude attenuation associated with the wave traveling from the first interface into the second one or vice versa. 
The semi-analytical model is able to account for the multiple resonances of the MMAs. Moreover, it is valid for \textit{TEM}, \textit{TE} and \textit{TM} polarizations and for a wide range of incidence angles.

It is noteworthy mentioning that the effective medium approximation is a valid model to be regarded for the 4-bit MMA structure, as the MMAs can be considered locally periodic. The unit-cell elements of the digitized MMA are designed to resonate at four discrete frequencies in the range of 71 GHz to 77 GHz, which is equivalent to a relatively low bandwidth (8.1\%). The scaling factor for the smallest and biggest unit-cells are $S=0.93$ and $S=1$, respectively, which means the largest geometry variation between the adjacent cells is only 7\%. Under these assumptions, it is safe to consider that the variation between the adjacent cells is smooth; therefore, local periodicity approach can be considered to establish an effective medium approximation model to predict the behavior of the digitized MMAs \cite{encinar2006dual,encinar2010recent}. 


\section{Imaging Metallic Scatterers using a 4-bit MMA-based CRA}

A 4-bit MMA-based CRA is used in an active mm-wave radar system to image metallic scatterers. The radar is composed of a CRA that is excited by an array of feeding elements. The parameters of the antenna elements in the feeding array, such as their quantity, spacing, and amplitude tapering through the surface of the reflector, are essential aspects of the imaging system. These parameters relate to the effective aperture of the feeding array, which defines the array factor in the far-field domain. The array factor is created by the physical apertures of the transmitters and receivers, and is governed by the double-convolution operation \cite{lockwood1994optimizing,ahmed2009near}. It can be inferred from this operation that for having an effective aperture area, it is not necessary to fill the whole area with radiating elements. Instead, by engineering the placement of the radiating elements, one can generate an effective aperture in the desired area. The feeding array parameters should be selected in a way that the effective aperture is able to cover the whole imaging region. Here, conical horn antennas are employed as the feeding elements. The transmitters and receivers are uniformly spaced in the $x$-axis and $y$-axis, respectively, forming a \textit{plus} configuration, as illustrated in Fig. \ref{Geo}. The physical and electrical parameters of the imaging system are described in Table \ref{tab:config1}.

Range and cross-range resolutions are the most important parameters determining imaging performance of any sensing system. Their limits for far-field imaging are, respectively, $R_r=\frac{c}{2BW}$ and $R_{cr}=\frac{\lambda_c R}{2D_{eff}}$ (both in $[m]$) \cite{byron1993radar,sleasman2016design}; in which $c$ is the velocity of light in $[m/s]$, $BW$ is the bandwidth of the radar in $[Hz]$, $\lambda_c$ is the wavelength in $[m]$ at the center frequency, $R$ is the distance between the physical aperture and the target in $[m]$, and $D_{eff}$ is the effective length of the physical aperture in $[m]$ in the direction where the beam-width is to be measured in $[m]$. However, for near-field imaging, the above-mentioned equations provide an estimation for the range and cross-range resolution and can be used to approximate the resolution performance of the system.

The measurements were collected through the $7$ \textit{GHz} frequency span with a center frequency of $73.5$ \textit{GHz}. The range distance between the focal plane of the CRA and the center of the imaging domain was $84$ \textit{cm}, and $D_{eff}$ in both $x$-axis and $y$-axis was $30$ \textit{cm}. Following the above-mentioned values, the range and cross-range resolution limits of the imaging system is calculated to be $R_r=21.4$ \textit{mm} and $R_{cr}=2.3$ \textit{mm}, respectively. Accordingly, the imaging domain was meshed to comply with the estimated range and cross-range resolution. 

An in-house numerical solver (MECA) was used for performing the simulations. The solver is based on physical optics and uses a modified equivalent current approximation \cite{Meana2010} to calculate the equivalent currents based on the oblique incidence of the plane wave on the interface. The solver discretizes the interfaces into planar triangular facets, on which induced currents are assumed to have constant magnitude and linear phase variation.

\begin{table}[htp]
\centering \caption{Design parameters of the 4-bit MMA-based imaging system.}\label{tab:config1}
\setlength{\extrarowheight}{1.5pt}
\begin{tabular}{|l|l|}
\hline
\multicolumn{1}{|c|}{\textbf{PARAMETER}} & \multicolumn{1}{c|}{\textbf{VALUE}} \\
\hline
\multicolumn{1}{|c|}{Frequency band} & \multicolumn{1}{c|}{$70-77$ \textit{GHz}} \\
\hline
\multicolumn{1}{|c|}{No. of frequencies ($N_f$)} & \multicolumn{1}{c|}{$15$} \\
\hline
\multicolumn{1}{|c|}{Reflector diameter ($D$)} & \multicolumn{1}{c|}{$30$ \textit{cm}} \\
\hline
\multicolumn{1}{|c|}{Focal length ($f$)} & \multicolumn{1}{c|}{$30$ \textit{cm}} \\
\hline
\multicolumn{1}{|c|}{No. of $Tx.$} & \multicolumn{1}{c|}{$4$} \\
\hline
\multicolumn{1}{|c|}{No. of $Rx.$} & \multicolumn{1}{c|}{$4$} \\
\hline
\multicolumn{1}{|c|}{Triangle size on the reflector} & \multicolumn{1}{c|}{$5\lambda_{c}=20.5$ \textit{mm}} \\
\hline
\multicolumn{1}{|c|}{ Number of measurements} & \multicolumn{1}{c|}{$N_{Tx}N_{Rx}N_{freq}=240$} \\
\hline
\multicolumn{1}{|c|}{ Number of pixels ($N_p$)} & \multicolumn{1}{c|}{$80,000$} \\
\hline
\multicolumn{1}{|c|}{ Range distance (${z_0}^T$)} & \multicolumn{1}{c|}{$84$ \textit{cm}} \\
\hline
\end{tabular}
\end{table}

\subsection{Sensing Capacity}

\begin{figure}[ht!]
    \centering
        \subfigure[]{\label{SVD}
        \includegraphics[width=0.45\textwidth,trim={0.2cm 1.5cm 1.2cm 3cm},clip=true]{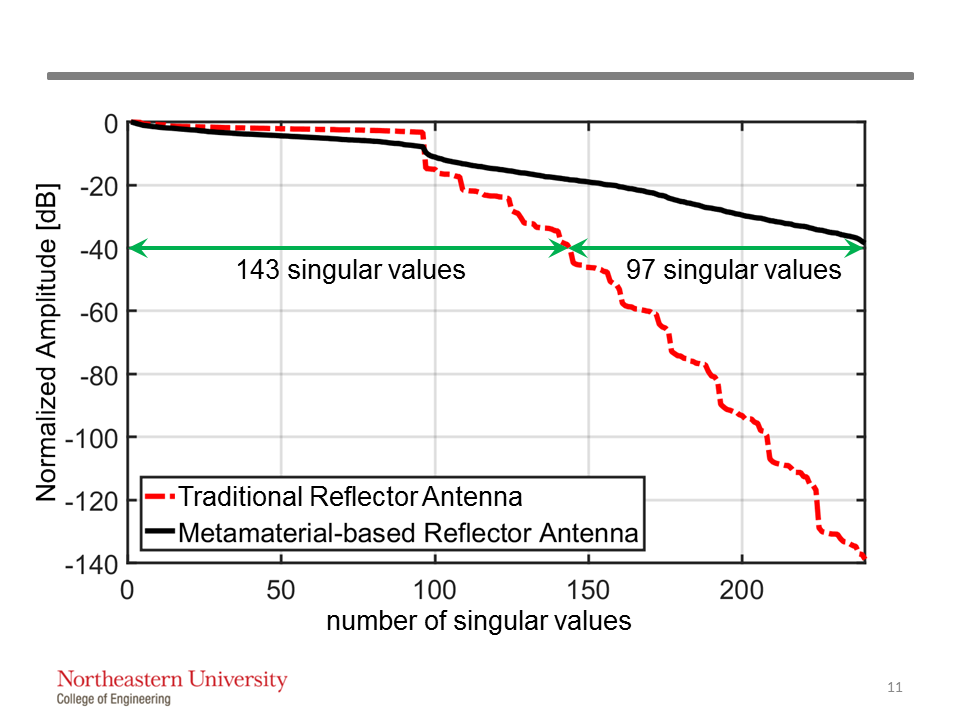}}\hspace{-0.5cm}
				
				\subfigure[]{\label{Capacity}
        \includegraphics[width=0.45\textwidth,trim={0.2cm 1.5cm 1.2cm 3cm},clip=true]{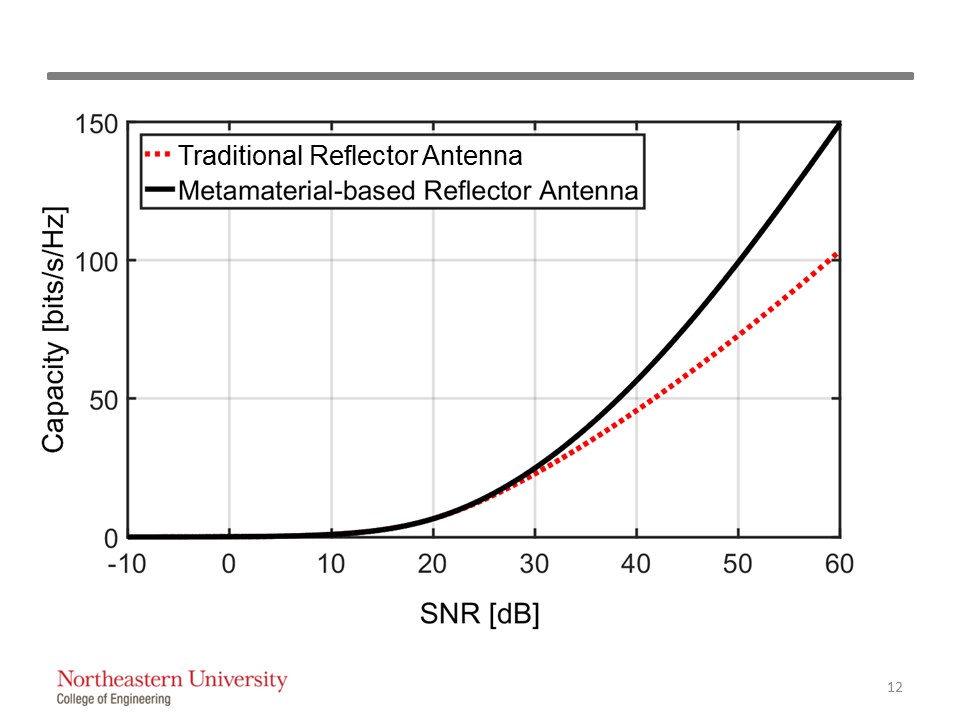}}

    \caption{ \label{SVD_Capacity} (a) Normalized SV distribution and (b) sensing capacity comparison between the TRA and the 4-bit CRA.}
\end{figure}

In order to obtain the SV distribution of the system, the sensing matrix $\mathbf{H}$ was calculated using MECA. Figure \ref{SVD} shows the improved SV distribution of the CRA when compared to that of the TRA, and Fig. \ref{Capacity} shows how the sensing capacity of the CRA is enhanced when the Signal to Noise Ratio (SNR) is increased. It is clear from Fig. \ref{Capacity} that by adding the MMAs, the sensing capacity of the imaging system improves. For example, if the sensing channel has an SNR level equal to $40$ \textit{dB}, the CRA would have $97$ more number of effective SVs above the noise floor, than the TRA. Ultimately, the sensing capacity of the CRA is improved, which can be inferred as an enhancement in the imaging performance.

\subsection{Radiation Pattern}

\begin{figure}[ht!]
    \centering
        \subfigure[]{\label{Pattern_a}
        \includegraphics[width=0.22\textwidth,trim={3.5cm 1.5cm 4cm 3cm},clip=true]{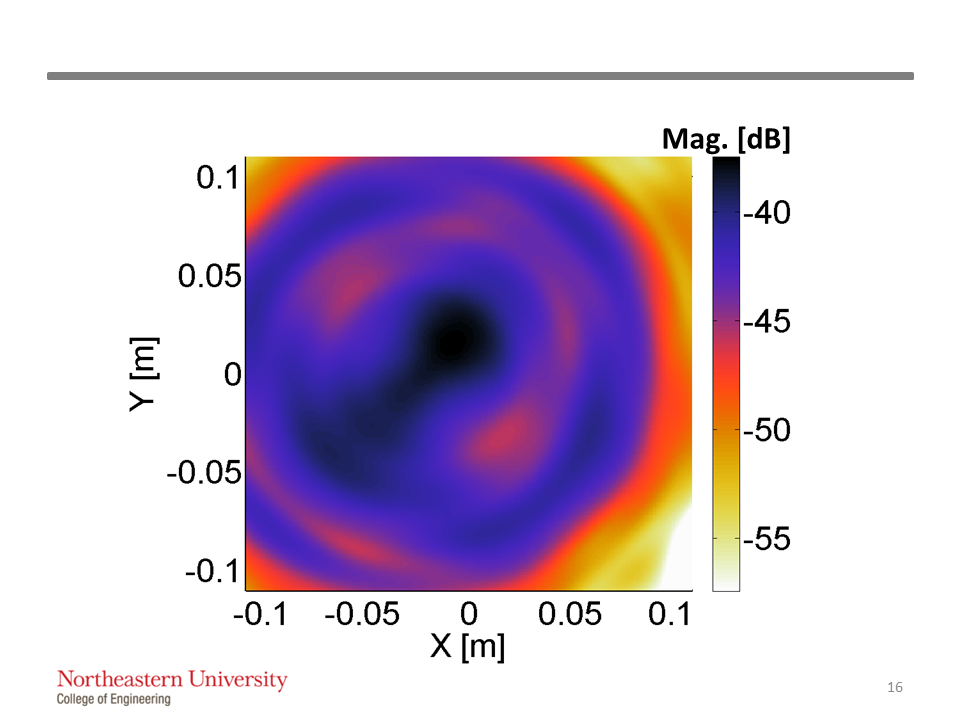}}
				~
				\subfigure[]{\label{Pattern_b}	
        \includegraphics[width=0.22\textwidth,trim={3.5cm 1.5cm 4cm 3cm},clip=true]{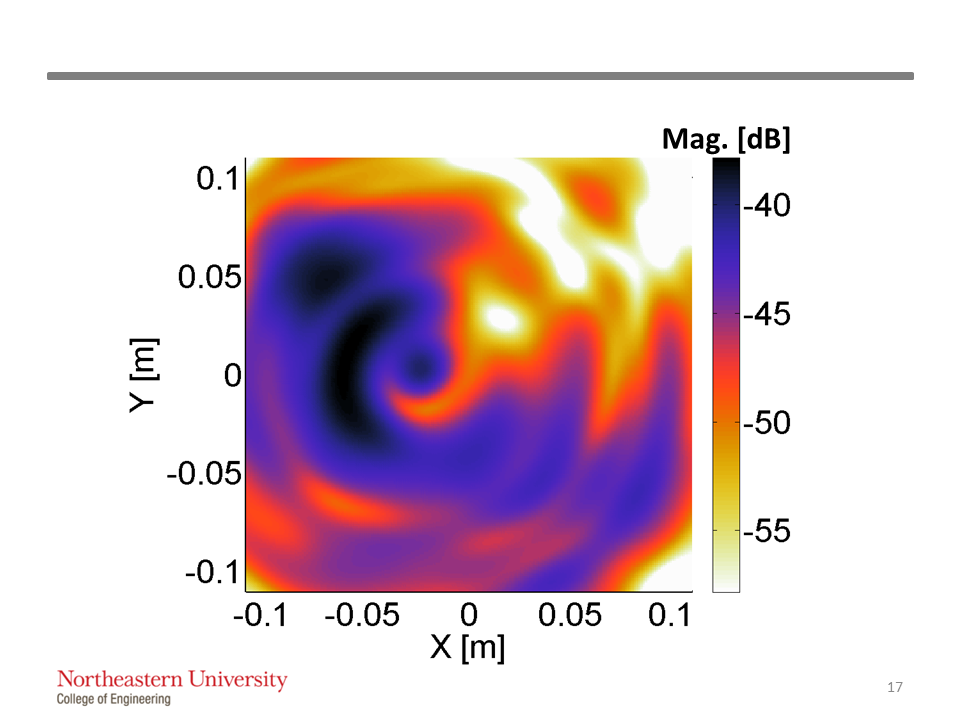}} 
        
				\subfigure[]{\label{Pattern_c}
        \includegraphics[width=0.22\textwidth,trim={3.5cm 1.5cm 4cm 3cm},clip=true]{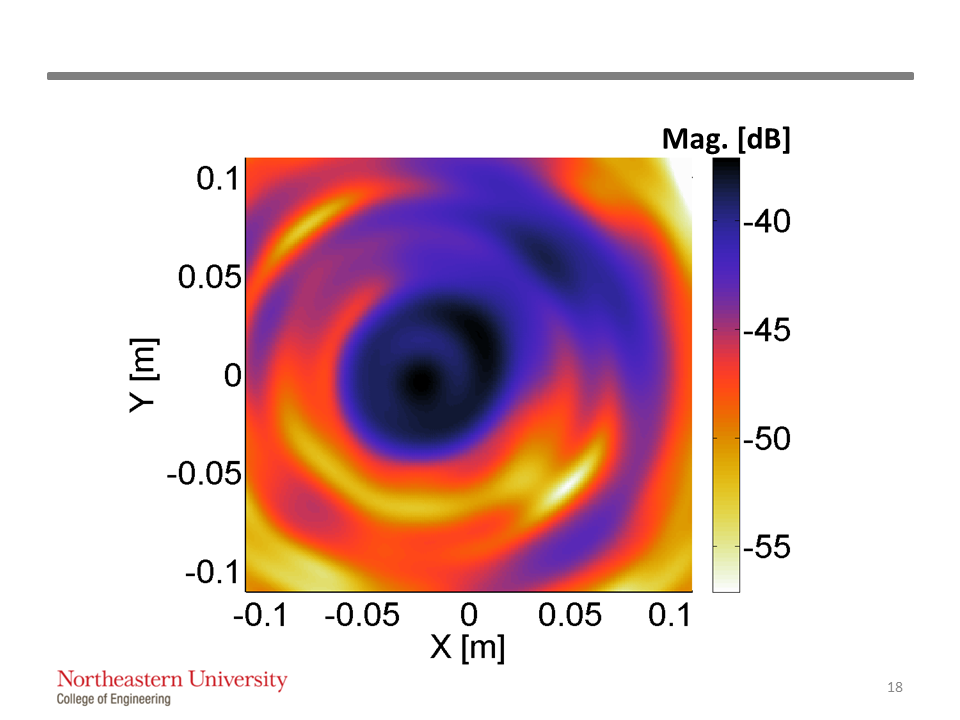}} 
				~			
				\subfigure[]{\label{Pattern_d}
        \includegraphics[width=0.22\textwidth,trim={3.5cm 1.5cm 4cm 3cm},clip=true]{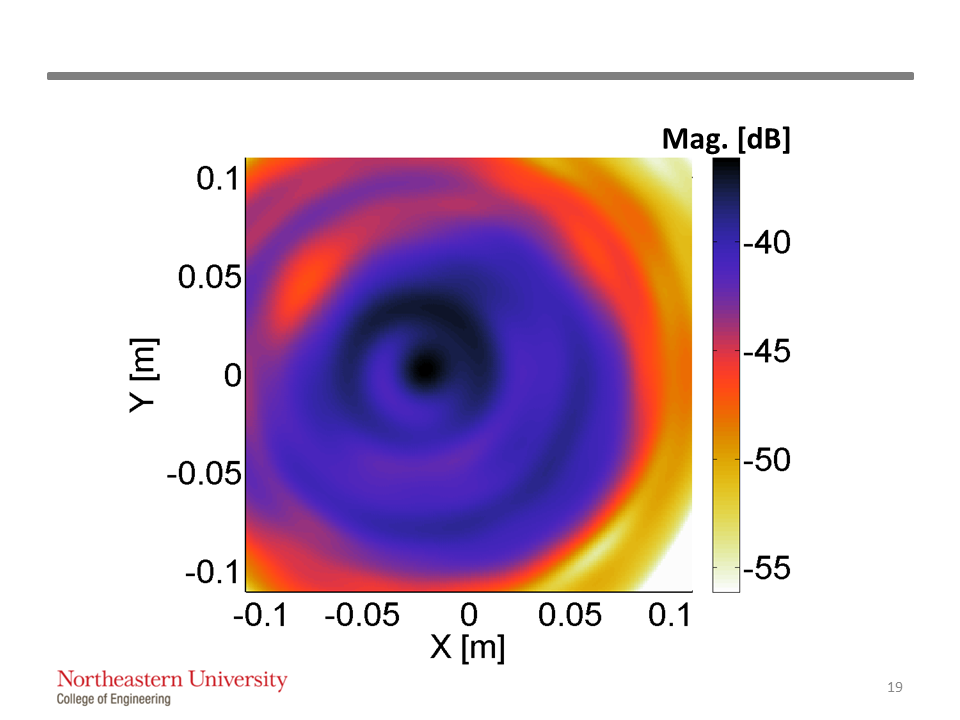}} 
        											
    \caption{ \label{Pattern} Electric field radiation pattern of the CRA system excited with $Rx_2$, plotted at the range of $85$ \textit{cm}, at (a) $70.5$ \textit{GHz}, (b) $72.5$ \textit{GHz}, (c) $74.5$ \textit{GHz}, and (d) $76.5$ \textit{GHz}.}
\end{figure}

By plotting the fields in the imaging region, one can study the radiation patterns as a function of frequency. Figure \ref{Pattern} shows the $y$-component of the electric field magnitude of the MMA-based reflector calculated on the $xy$-plane, at a distance $z=85$ \textit{cm} from the focal plane. These radiation patterns correspond to the case where $Rx_2$ is feeding the reflector and they are plotted for $70.5$ \textit{GHz} [Fig. \ref{Pattern_a}], $72.5$ \textit{GHz} [Fig. \ref{Pattern_b}], $74.5$ \textit{GHz} [Fig. \ref{Pattern_c}], and $76.5$ \textit{GHz} [Fig. \ref{Pattern_d}]. As expected, the radiation pattern for each of the antennas varies as frequency changes. This behavior is due to the peculiar frequency dispersive response of the 4-bit MMAs.

\subsection{Beam Focusing}

\begin{figure}[ht!]
    \centering
        \subfigure[]{\label{Rx}
        \includegraphics[width=0.45\textwidth,trim={0.2cm 4cm 0.5cm 4cm},clip=true]{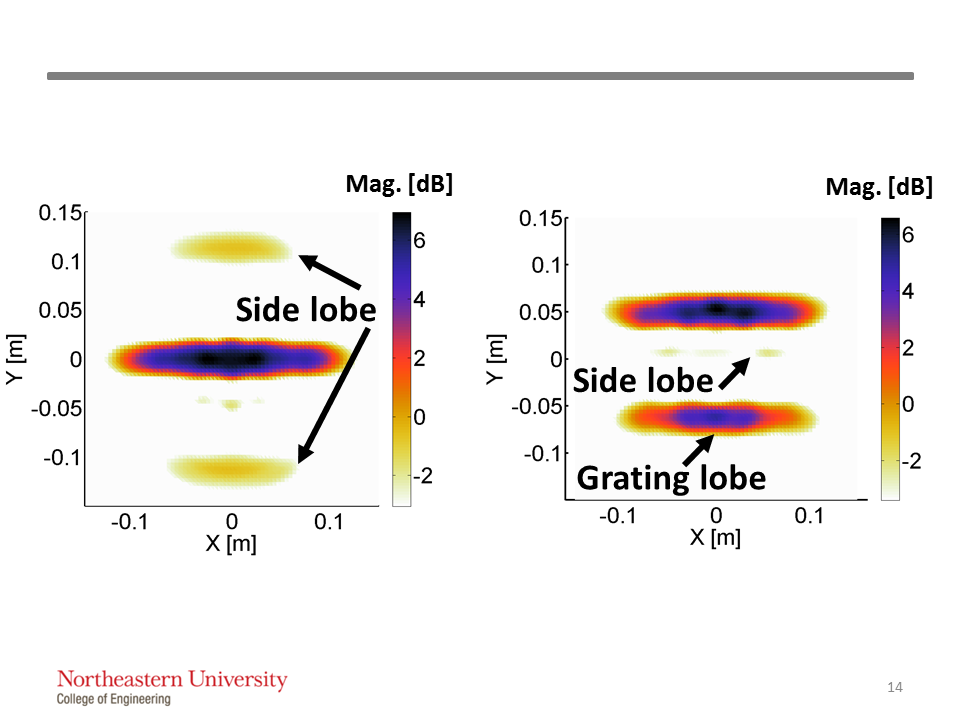}}

				\subfigure[]{\label{Tx}
        \includegraphics[width=0.45\textwidth,trim={0.2cm 4cm 0.5cm 4cm},clip=true]{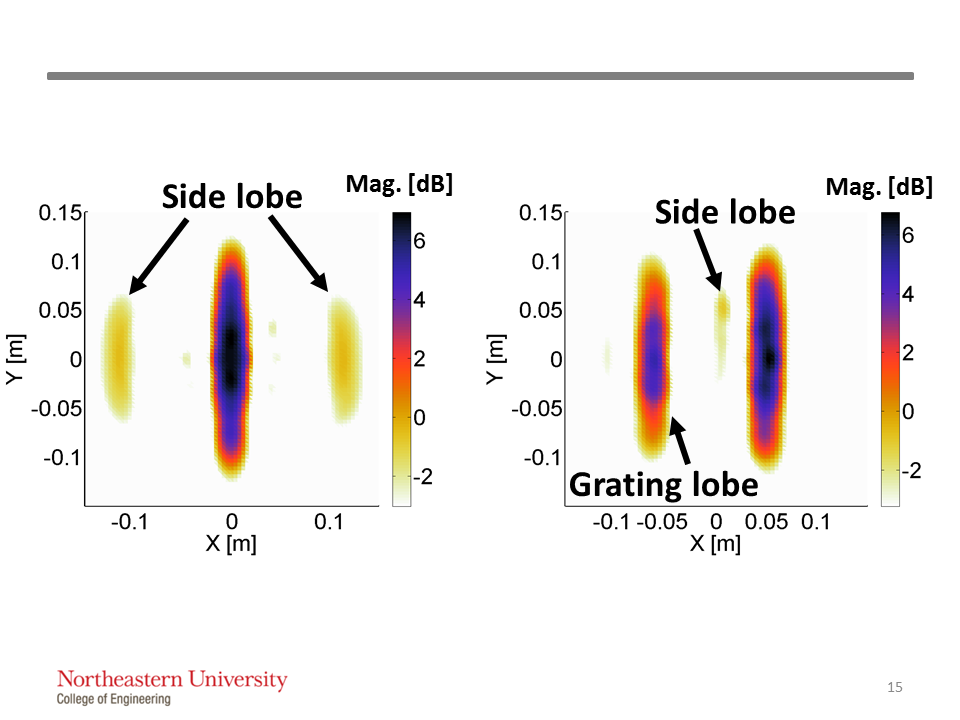}}

				\subfigure[]{\label{Mono}
        \includegraphics[width=0.47\textwidth,trim={0.2cm 4cm 0.5cm 4cm},clip=true]{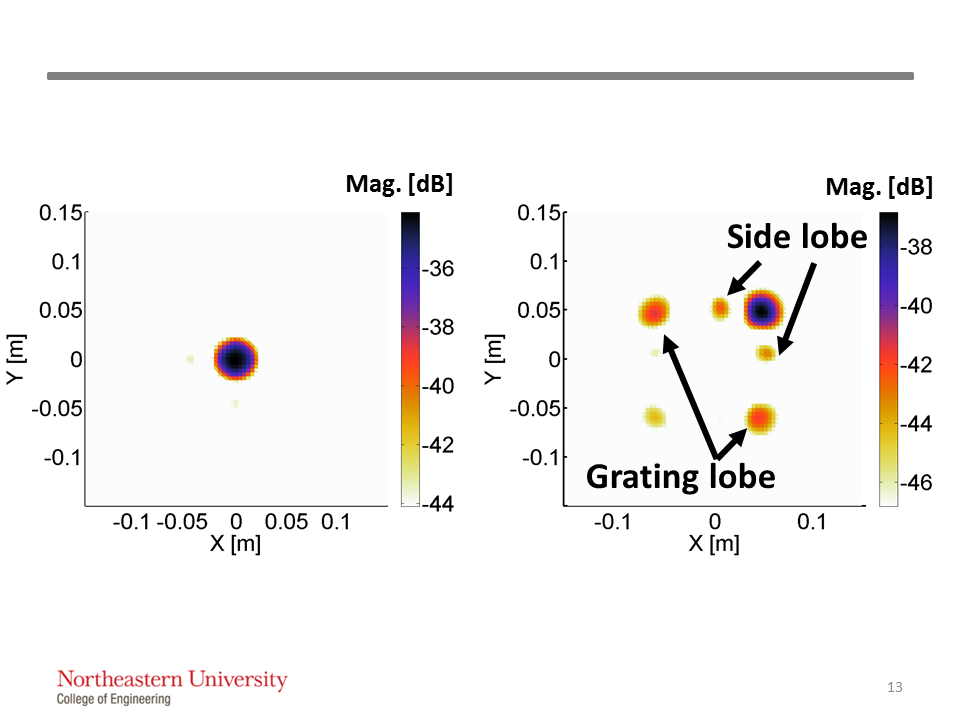}}

    \caption{ \label{PSF} Beam focusing at (left) $[x,y,z]=[0,0,85]cm$, and (right) $[x,y,z]=[5,5,85]cm$, by the (a) receiver array, (b) transmitter array, (c) and the product of the receiver and the transmitter array BFs.}
\end{figure}

To have a better understanding of the CRA performance, the Beam Focusing (BF) capability of the system is examined. The BF can be evaluated by applying a phase compensation method \cite{ahmed2009multistatic}. To accomplish this, for each measurement mode, a phase correction term is introduced to the \textit{Tx} and \textit{Rx} elements, to produce a phase equal to zero at the desired focusing point. Therefore, adding all the measurements results in a constructive summation at the focusing point. Such a study in the imaging region visualizes the resolution, side-lobes, grating lobes, and artifacts of the MMA-based reflector and helps to further understand the behavior of the imaging system. In this regard, focusing in two points ($[0,0,85]cm$ and $[5,5,85]cm$) of the observation domain is studied. Figure \ref{PSF} illustrates the obtained images of the BF, calculated using the phase compensation method. Figs. \ref{Rx} and \ref{Tx} show the BF of the receiver and transmitter array, respectively, and Fig. \ref{Mono} shows the product of the receiver and the transmitter array BFs. It is important to note here that the norm-1 regularized imaging algorithm described below is capable of overcoming the degrading effect of the grating lobes.


\subsection{Image Reconstruction}
\subsubsection{ADMM algorithm}

In order to accelerate the imaging process, the sensing matrix $\bf{H}$ in Eq. \ref{sensing_eq} may be split into $N$ sub-matrices $\bf{H}_i$ by rows, and the vector $\bf{g}$ into $N$ sub-vectors $\bf{g}_i$. Additionally, $N$ different variables $\bf{u}_i$ may be defined in order to solve $N$ independent sub-problems. By adding a constraint that enforces the \textit{consensus} of all solutions, the problem to optimize becomes the following:
\begin{equation}
\left.
\begin{array}{cc}
\mbox{minimize } & \frac{1}{2}\sum\limits_{i=1}^{N}\left\Vert
\textbf{H}_{i}\textbf{u}_i-\textbf{g}_{i}\right\Vert _{2}^{2}+\lambda\left\Vert \textbf{v}\right\Vert _{1} \\
\mbox{s.t.} & \textbf{u}_i=\textbf{v},\;\;\forall i=1,...,N.%
\end{array}%
\right.
\label{sum_N}
\end{equation}%


The variable $\bf{v}$ serves as a \textit{consensus} variable, imposing
the agreement between all the variables $\textbf{u}_i$. See for
example \cite{HerediaJuesas2015, degroot1974reaching,erseghe2011fast, mota2012distributed, mota2013d, forero2010consensus}.
This problem can be solved by the following iterative scheme:
\begin{eqnarray}
\label{x_solution}
\textbf{u}_{i}^{k+1} &=&\left( \textbf{H}_{i}^{\ast }\textbf{H}_{i}+\rho \textbf{I}\right) ^{-1}\left(\textbf{H}_{i}^{\ast }\textbf{g}_{i}+\rho \left(\textbf{v}^{k}-\textbf{s}_{i}^{k}\right) \right) , \\
\textbf{v}^{k+1} &=&\mathbf{S}_{\frac{\lambda}{\rho N}}\left( \bar{\textbf{u}}^{k+1}+\bar{\textbf{s}}^{k}\right) , \label{v_variable}\\
\textbf{s}_{i}^{k+1} &=&\textbf{s}_{i}^{k}+\textbf{u}_{i}^{k+1}-\textbf{v}^{k+1} ,
\end{eqnarray}%
where $\textbf{s}_i$ is the dual variable for the $i$-th constraint, $\rho $ is the augmented parameter
, and $\mathbf{S}_{\kappa }\left( \cdot \right) $ is the soft thresholding
operator \cite{bredies2008linear}
interpreted elementwise.
Moreover, $\bar{\textbf{u}}$ and $\bar{\textbf{s}}$
are the mean of $\textbf{u}_{i}$ and $\textbf{s}_{i}$, respectively, for all $i$. 
The term $\left( \textbf{H}_{i}^{\ast }\textbf{H}_{i}+\rho \textbf{I}\right) ^{-1}$ requires the inversion of an $N_p\times N_p$ matrix, which is computationally expensive. However,
the \textit{matrix inversion lemma} \cite{woodbury1950inverting} can be applied in order to perform $N$ inversions of matrices of reduced size $\frac{N_m}{N}\times \frac{N_m}{N}$, as equation \eqref{inversion_lemma} shows:
\begin{equation}
\label{inversion_lemma}
\left( \textbf{H}_{i}^{\ast }\textbf{H}_{i}+\rho \textbf{I}_{N_p}\right) ^{-1}=\frac{\textbf{I}_{N_p}}{\rho }-\frac{%
\textbf{H}_{i}^{\ast }}{\rho ^{2}}\left( \textbf{I}_{\frac{N_m}{N}}+\frac{\textbf{H}_{i}\textbf{H}_{i}^{\ast }}{\rho }%
\right) ^{-1}\textbf{H}_{i},
\end{equation}
where $\textbf{I}_{N_p}$ and $\textbf{I}_{\frac{N_m}{N}}$ indicate the identity matrices of sizes $N_p$ and ${\frac{N_m}{N}}$, respectively. Since the consensus-based ADMM algorithm is efficiently parallelizable, it can be solved by running a MATLAB code using a GPU, reducing the computational time of the imaging process.

\subsubsection{Imaging results}

\begin{figure}[ht!]
    \centering
        \subfigure[]{\label{Image_NoMMA}
        \includegraphics[width=0.5\textwidth,trim={4.1cm 1.2cm 5cm 4cm},clip=true]{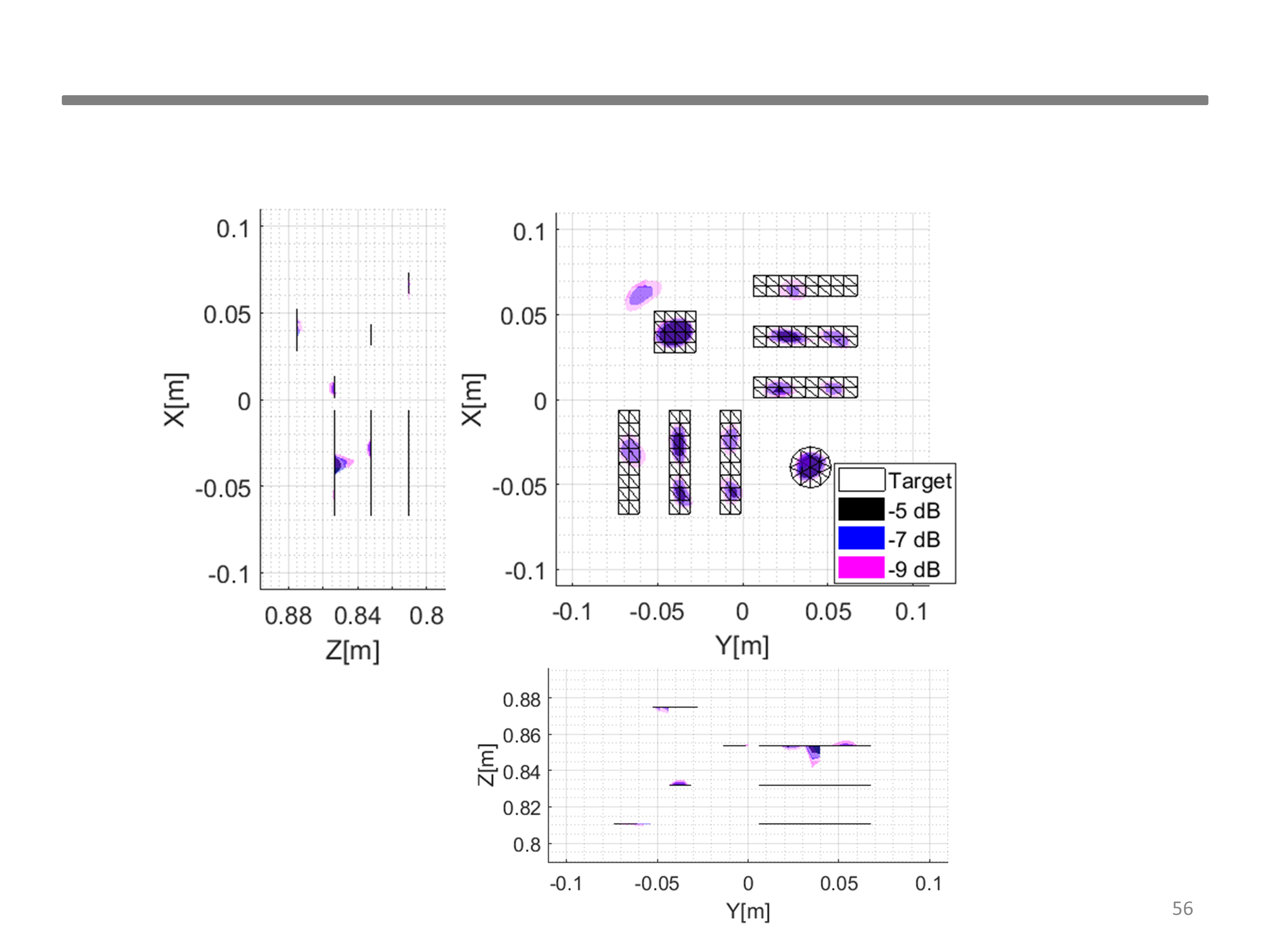}}\hspace{-0.3cm}

				\subfigure[]{\label{Image_MMA}
        \includegraphics[width=0.5\textwidth,trim={4.1cm 1.5cm 5cm 4cm},clip=true]{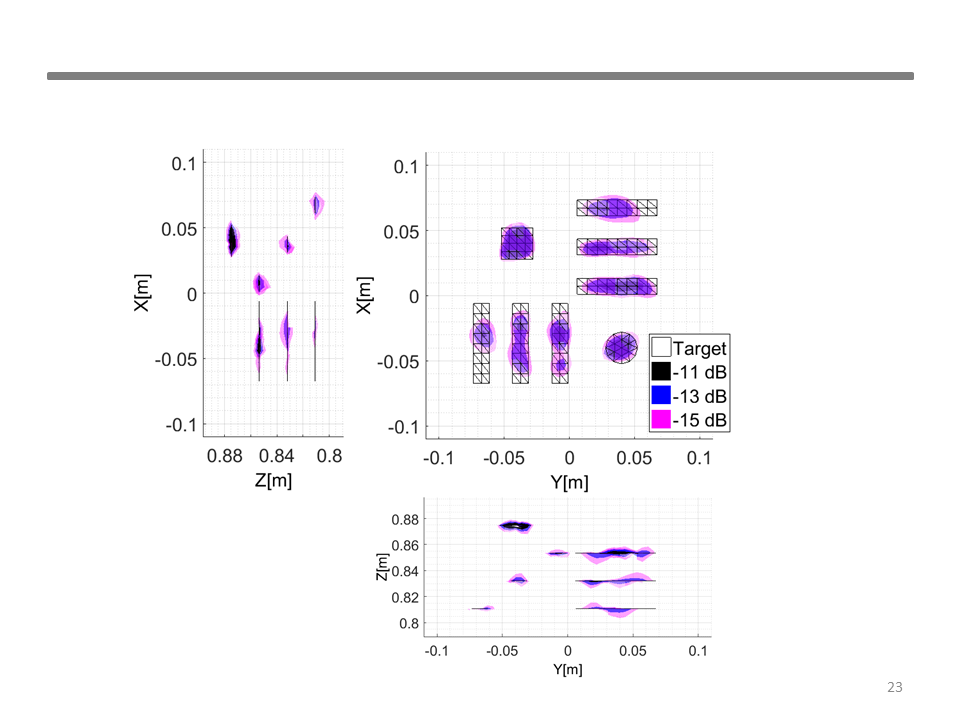}}

    \caption{ \label{ADMM} Reconstructed image using ADMM iterative method for (a) the TRA and (b) the 4-bit CRA. }
\end{figure}

In this example, 2D planar PEC objects located at different ranges are considered as the targets. 
The objects are three vertically and three horizontally oriented rectangles, one square, and one circle. The width and height of the rectangles are equal to $12$ \textit{mm} and $61$ \textit{mm}, respectively. The side of the square, as well as the diameter of the circle, is $24$ \textit{mm}. The center of the imaging domain is considered to be $84$ \textit{cm} far from from the focal plane of the reflector and the PEC scatterers are located in four different range positions, separated by $21$ \textit{mm}. The reconstruction for imaging using the TRA and CRA are shown in Figs. \ref{Image_NoMMA} and \ref{Image_MMA}, respectively. The imaging has been carried out through the ADMM algorithm. The regularization parameters of ADMM, which were optimized for each imaging case separately, are as follows: $\rho=0.1$, $\lambda=5$, and $N=48$ for the TRA; and $\rho=1$, $\lambda=100$, and $N=12$ for the CRA. It is evident that the CRA has an improved imaging, compared to that of the TRA.

\section{Imaging a Human Model Using an Array of CRAs}

\begin{figure}[H]
    \centering
        \subfigure[]{\label{CRA_Array}
        \includegraphics[width=0.5\textwidth,trim={0cm 0cm 1.7cm 5cm},clip=true]{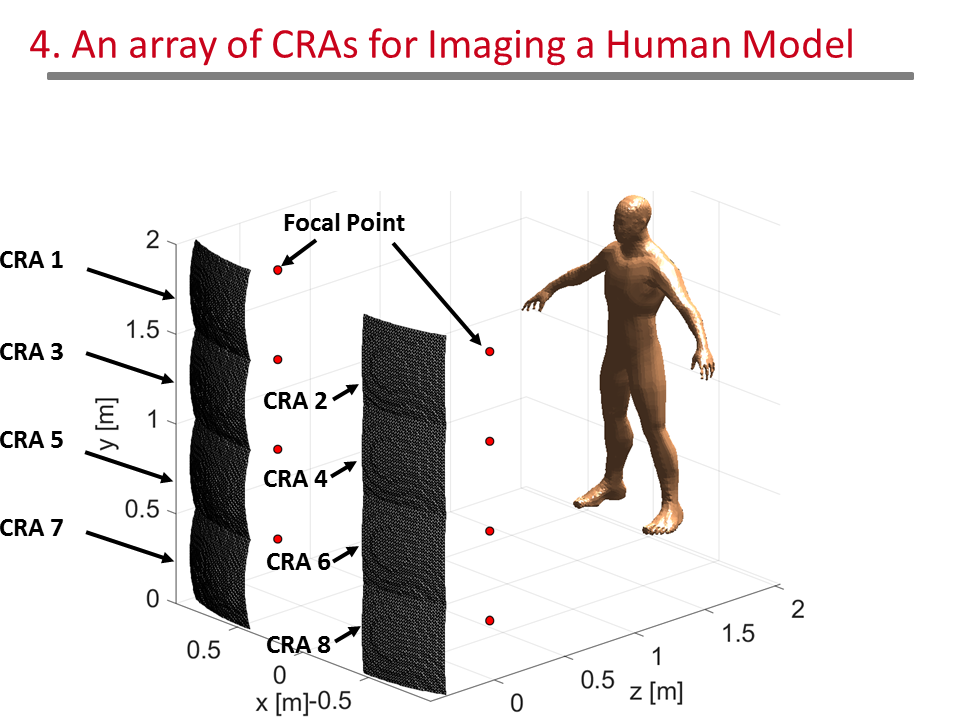}}

				\subfigure[]{\label{Single_CRA}
        \includegraphics[width=0.45\textwidth,trim={0cm 0cm 0cm 5cm},clip=true]{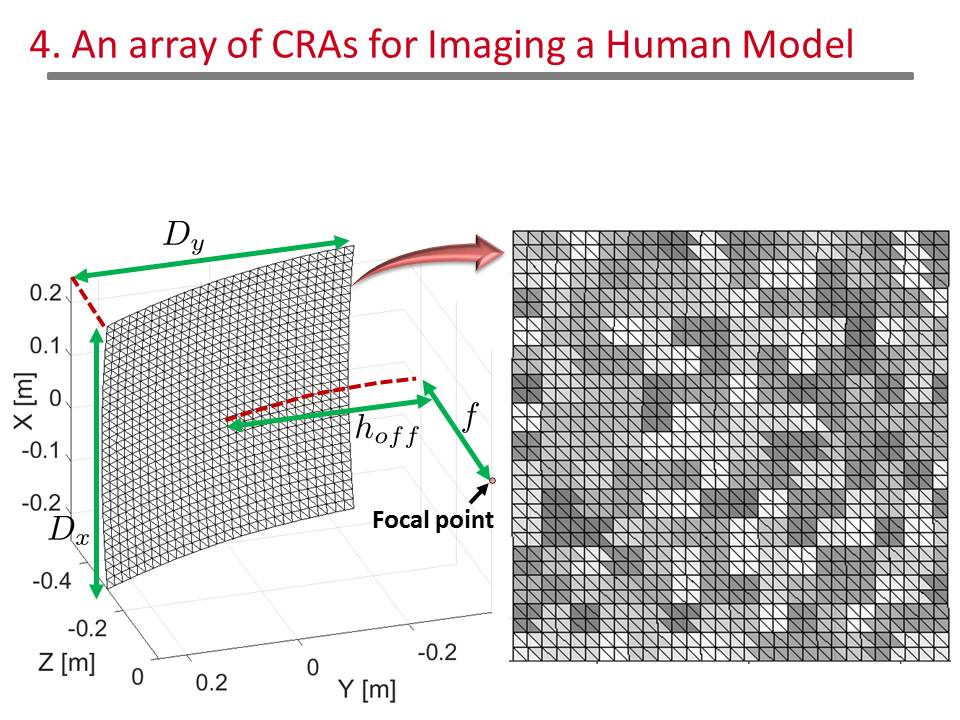}} 
    \caption{ \label{2nd_Example} (a) CRA array for imaging a human model; (b) Left: a single CRA in offset mode. Right: the 8-bit MMAs are tailored on the surface of the CRA.}
\end{figure}

As another example of digitized MMAs, a reflector-based array made of eight CRAs is used to image an extended human-size region, as shown in Fig. \ref{CRA_Array}. The 3D human model was projected into a 2D plane located $2 \; m$ away from the focal plane. The coding mechanism of the CRAs relies on two principles: 1) spatial coding of the electromagnetic field, generated by introducing discrete PEC scatterers on the surface of a TRA--see \cite{Martinez-Lorenzo2015} for a detailed explanation on spatially coded CRAs--and 2) spectral coding of electromagnetic fields, generated by 8-bit MMAs, tailored on the surface of the reflector.

\begin{figure}[H]
    \centering
        \subfigure[]{\label{Single_CRA_Rx}
        \includegraphics[width=0.22\textwidth,trim={0cm 0cm 13cm 6.5cm},clip=true]{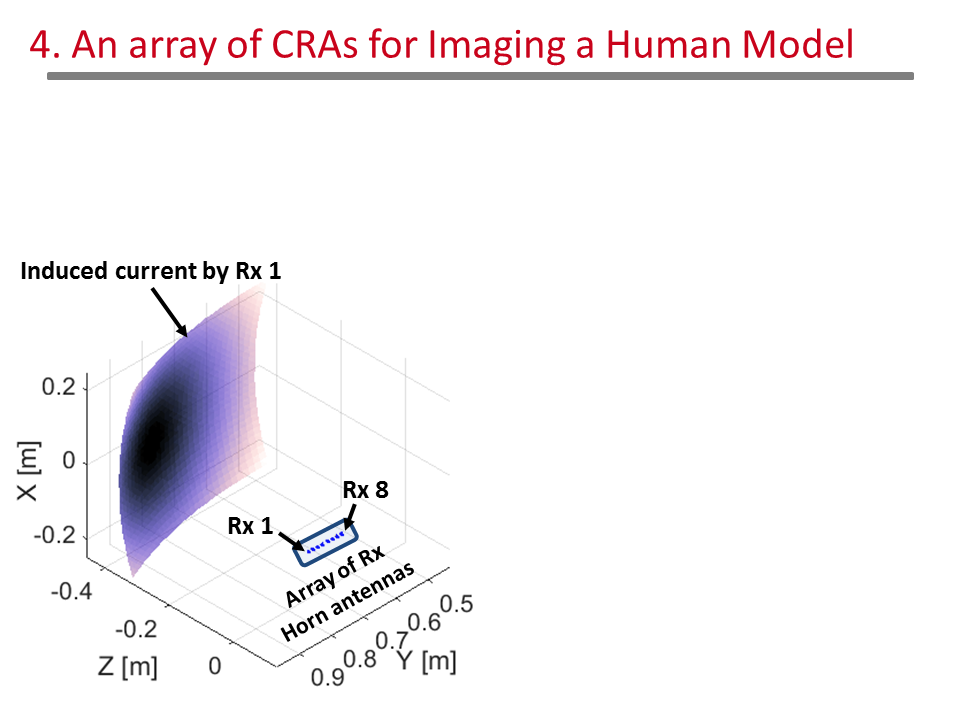}}
        ~
				\subfigure[]{\label{Single_CRA_Tx}
        \includegraphics[width=0.22\textwidth,trim={0cm 0cm 13cm 6.5cm},clip=true]{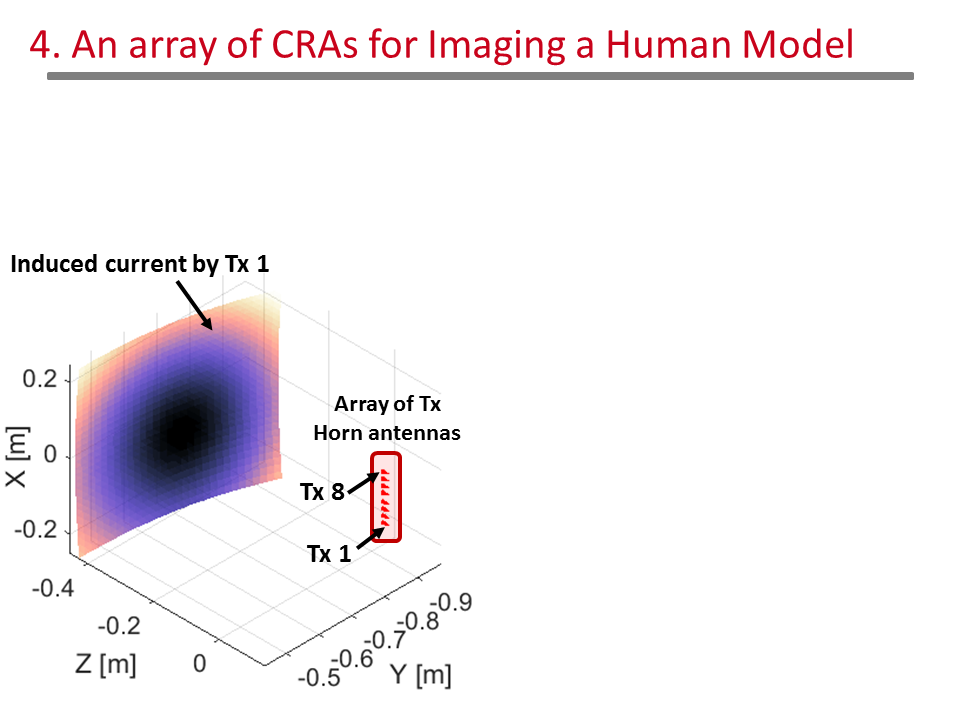}} 
    \caption{ \label{Single_CRA_feed} (a) Array of receiving horn antennas feeding the CRA-$2$ (b) array of transmitting horn antennas feeding the CRA-$1$.}
\end{figure}

Each CRA is illuminated with two orthogonal transmitting [Fig. \ref{Single_CRA_Tx}] and receiving [Fig. \ref{Single_CRA_Rx}] arrays located on the focal plane of the reflector. The electromagnetic cross-coupling between adjacent CRAs was used to perform the imaging. Given the location of the target with respect to the array, only the electromagnetic cross-coupling between CRA-{\it l} and CRA-{\it k}: ($l=1,k=2$), ($l=3,k=4$), ($l=5,k=6$), ($l=7,k=8$), ($l=1,k=4$), ($l=3,k=6$) and ($l=5,k=8$) is considered.

\begin{figure}[H]
    \centering
        \includegraphics[width=0.5\textwidth,trim={0cm 0cm 0cm 0cm},clip=true]{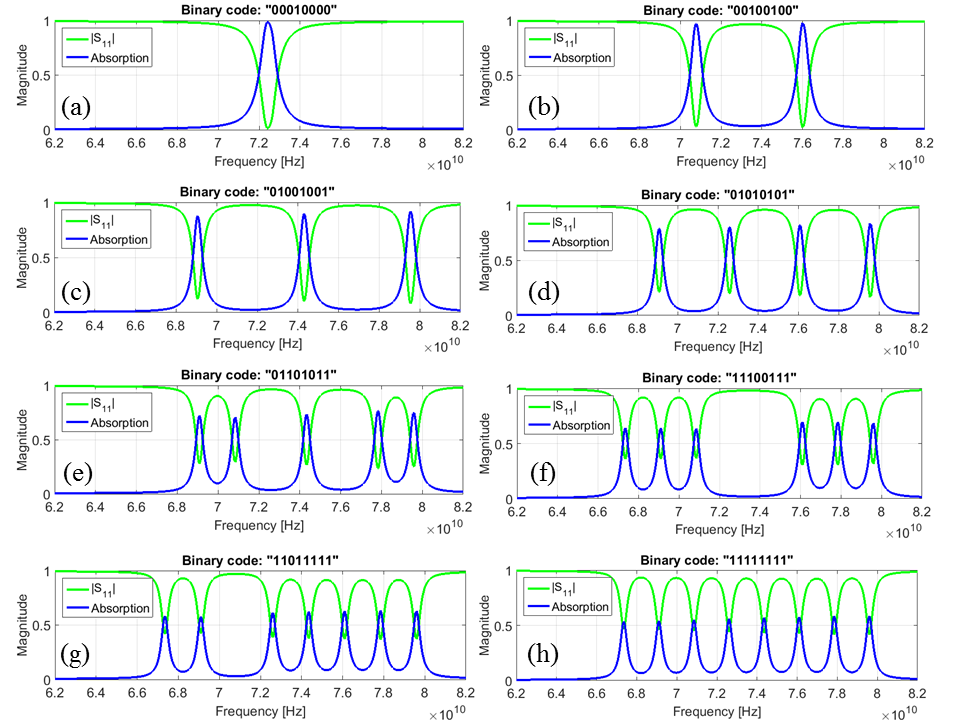}\vspace{-5pt}
    \caption{ \label{8DigitsCode} Magnitude of reflection ($|S_{11}|$) and absorption ($1-|S_{11}|^2$) of the MMA array for different binary codes: (a) $c_{16}=00010000$, (b) $c_{32}=00100100$, (c) $c_{73}=01001001$, (d) $c_{85}=01010101$, (e) $c_{107}=01101011$, (f) $c_{231}=11100111$, (g) $c_{223}=11011111$, and (h) $c_{255}=11111111$.}
\end{figure}

The design parameters for each one of the reflectors are shown in Table \ref{Tab_Single_CRA}--see \cite{Martinez-Lorenzo2015} for a detailed explanation on the meaning of these parameters. Both the vertical receiving array [Fig. \ref{Single_CRA_Rx}] and the horizontal transmitting array [Fig. \ref{Single_CRA_Tx}] of each CRA consist of $8$ uniformly distributed conical horn antennas. With a similar procedure explained in Section \ref{Sec_MMA}, an 8-bit binary code $c_i=a_1...a_8$ ($i \in \{ 1,...,256 \}$) is associated to each MMA design. Each digit of the binary code $a_j \in \{0,1\}$ ($j \in \{1,2,...,8\}$) is associated with a resonance frequency $f_j$. The eight resonant frequencies are uniformly selected within the $67.375$ GHz to $79.625$ GHz frequency band. The reflector's surface is divided into $2^8=256$ sub-surfaces and is randomly coated with the $256$ different MMA configurations. In Fig. \ref{Single_CRA}, each gray-scale color ($rgb(i/2,i/2,i/2)$) on the surface of the reflector represents a unique MMA configuration operating with the code $c_i$. In Fig. \ref{8DigitsCode}, the reflection coefficient and absorption value for eight different binary codes are plotted. Based on the resonance frequencies of the 8-bit MMAs, $66.5-80.5$ \textit{GHz} is selected to be the operational band of the radar and $N_f=87$ regularly sampled frequencies are used to perform the imaging. 

In this numerical example, the 8-bit MMAs are established in the frequency range of 66.5 GHz to 80.5 GHz, equivalent to a bandwidth of 19\%, which is more than twice the bandwidth of the 4-bit MMA example presented in the previous example. Nevertheless, there have been several efforts on extending the local periodicity approach \cite{zhou2011analysis}, which aim at expanding application scope of the method to the cases where the variations between the elements are not smooth. As it can be seen from the results in \cite{zhou2011analysis}, in such cases the deviations and inaccuracies predominantly occur at the side lobes while the predictions of local periodicity for the main lobe remains accurate. In the case of this example, there is not much concern about the side lobes and they do not play a major role in the near-field sensing, which makes effective medium approach valid, even if the variations between the adjacent elements are not very smooth.

\begin{table}[htp]
\centering \caption{Design parameters for a single CRA.}\label{Tab_Single_CRA}
\setlength{\extrarowheight}{1.5pt}
\begin{tabular}{|l|l|}
\hline
\multicolumn{1}{|c|}{\textbf{PARAMETER}} & \multicolumn{1}{c|}{\textbf{VALUE}} \\
\hline
\multicolumn{1}{|c|}{Frequency band} & \multicolumn{1}{c|}{$66.5-80.5$ $GHz$} \\
\hline
\multicolumn{1}{|c|}{No. of frequencies ($N_f$)} & \multicolumn{1}{c|}{$87$} \\
\hline
\multicolumn{1}{|c|}{No. of $Tx.$} & \multicolumn{1}{c|}{$8$} \\
\hline
\multicolumn{1}{|c|}{No. of $Rx.$} & \multicolumn{1}{c|}{$8$} \\
\hline
\multicolumn{1}{|c|}{Aperture size $(D_x=D_y$)} & \multicolumn{1}{c|}{$50$ $cm$} \\
\hline
\multicolumn{1}{|c|}{Focal length ($f$)} & \multicolumn{1}{c|}{$50$ $cm$} \\
\hline
\multicolumn{1}{|c|}{Offset height ($h_{off}$)} & \multicolumn{1}{c|}{$35$ $cm$} \\
\hline
\multicolumn{1}{|c|}{Size of the random facets} & \multicolumn{1}{c|}{$4\lambda$} \\
\hline
\multicolumn{1}{|c|}{Max. Distortion of facets} & \multicolumn{1}{c|}{$3^{\circ}$} \\
\hline
\end{tabular}
\end{table}

\begin{figure}[H]
    \centering
        \subfigure[]{\label{HM_SVD}
        \includegraphics[width=0.35\textwidth,trim={0cm 0cm 0cm 4cm},clip=true]{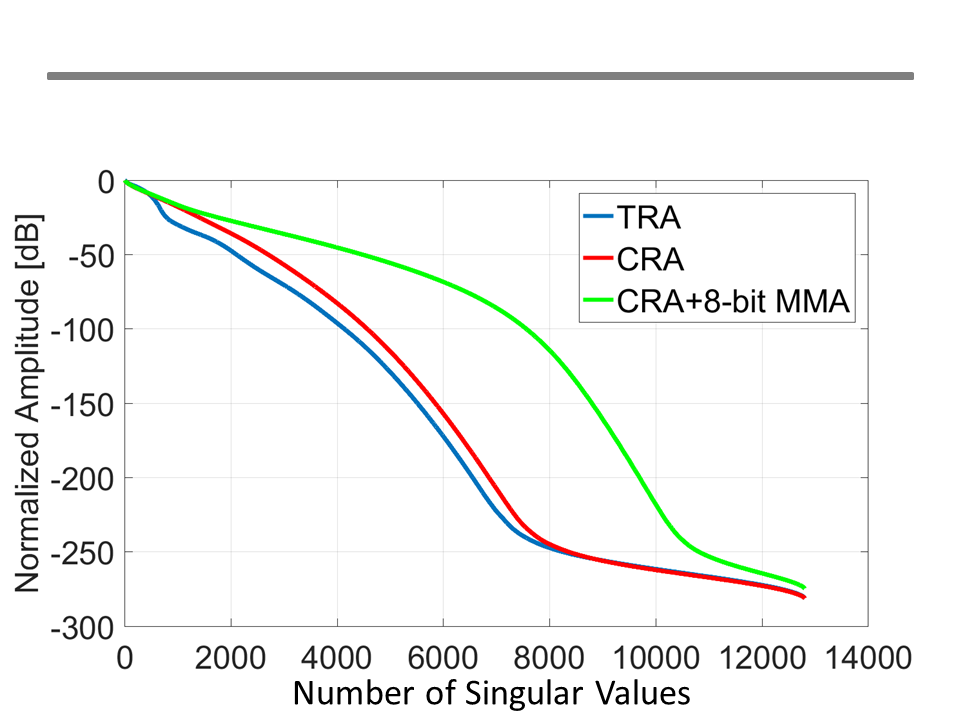}}\vspace{-0.18cm}
        ~
				\subfigure[]{\label{HM_Capacity}
        \includegraphics[width=0.35\textwidth,trim={0cm 0cm 0cm 4cm},clip=true]{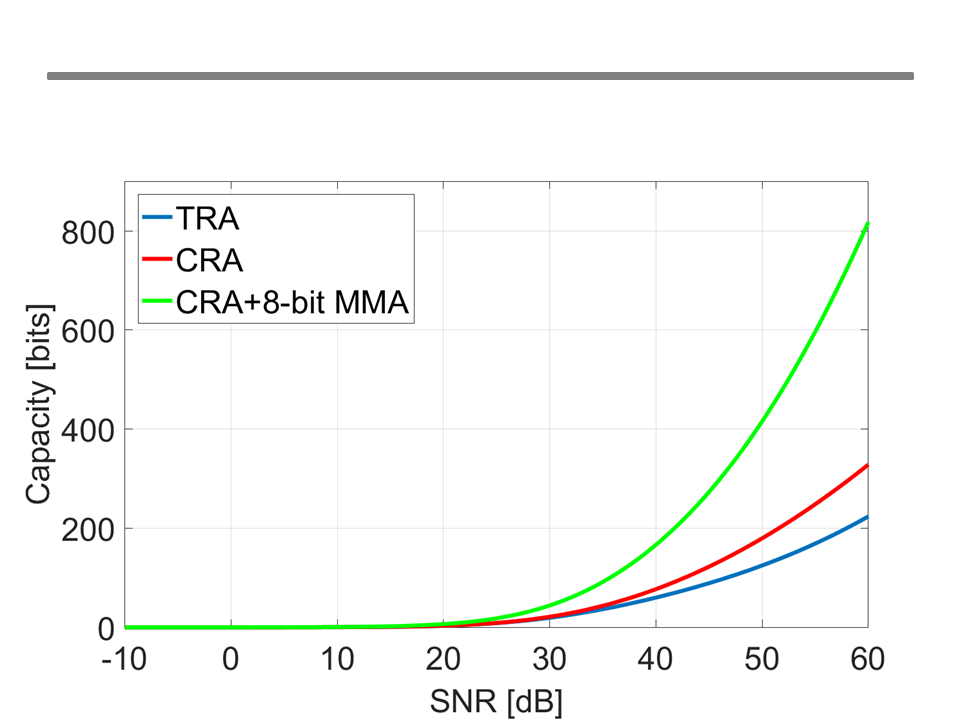}} 
    \caption{ \label{Human_Model_SVD_Capacity} Comparison of (a) the normalized SV distribution and (b) the sensing capacity of a single CRA and TRA.}\vspace{-0.18cm}
\end{figure}

\begin{figure}[H]
	\centering
	\subfigure[]{\label{HM_NESTA_NoMMA}
		\includegraphics[width=0.28\textwidth,trim={0cm 0cm 9cm 0cm},clip=true]{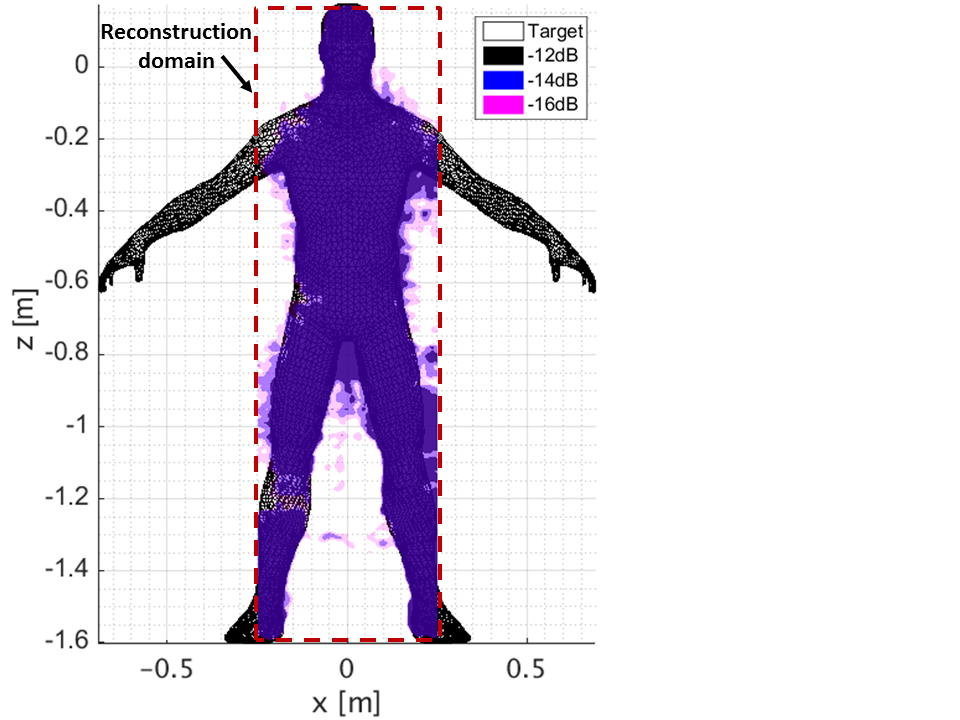}}
	~
	\subfigure[]{\label{HM_NESTA_MMA}
		\includegraphics[width=0.28\textwidth,trim={0cm 0cm 9cm 0cm},clip=true]{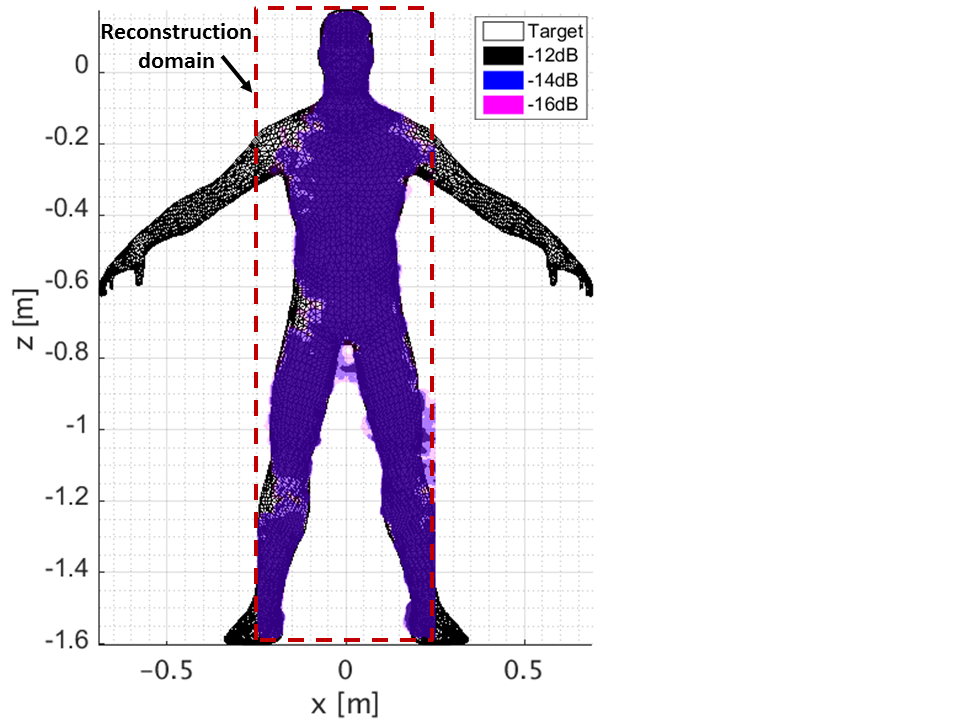}} 
	\caption{ \label{HM_NESTA} Reconstructed image using iterative compressive sensing algorithm (NESTA) for the (a) CRA and (b) MMA-based CRA arrays.}
\end{figure}

Figure \ref{HM_SVD} shows the improved SV distribution of the single MMA-based CRA, when compared to that of the CRA without MMAs and the TRA. Also, Fig. \ref{HM_Capacity} shows how the sensing capacity of the CRA is enhanced for different SNR levels. The image reconstruction is performed using the MATLAB toolbox NESTA \cite{Becker2011} imposing norm-1 regularization priors. The imaging result of the CRA with and without the 8-bit MMAs are plotted in Fig. \ref{HM_NESTA}.

\begin{figure}[htp]
	\centering
	\includegraphics[width=0.48\textwidth,trim={0cm 0cm 6cm 8cm},clip=true]{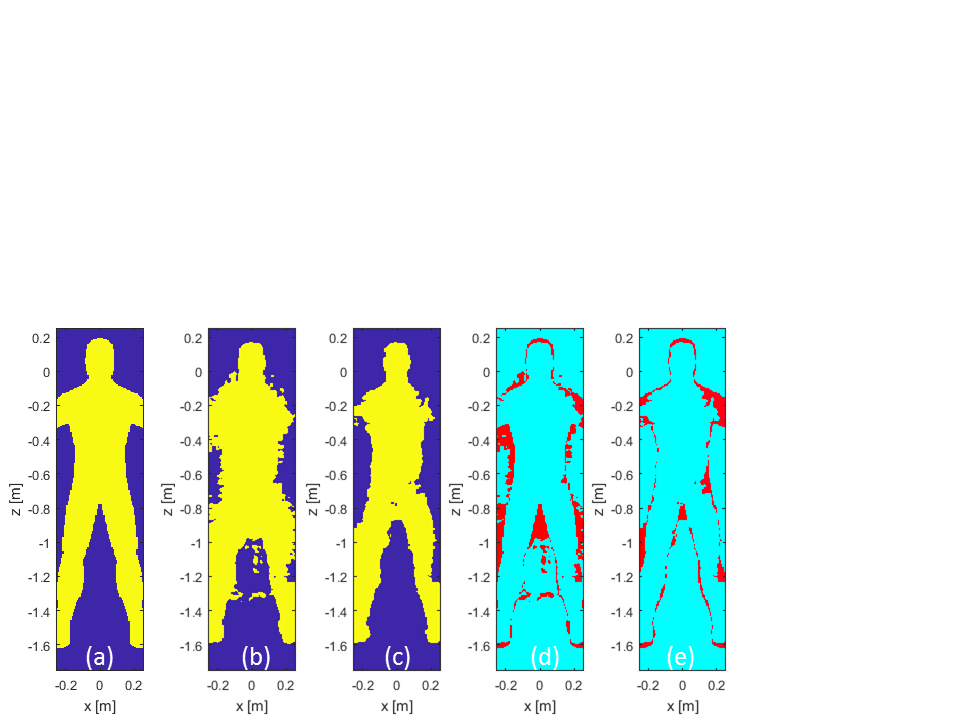}
	\caption{(a) Original target. Reconstructed image above the -16 dB threshold for the (b) CRA and (c) MMA-based CRA configurations; Image reconstruction error (red color) for the (d) CRA and (e) MMA-based CRA configurations.}
	\label{error}
\end{figure}

To justify the improvement of the MMA-based CRA compared to that of the CRA without the MMAs, the imaging accuracy for both configurations are quantitatively measured and then compared to each other. Fig. \ref{error}(a) shows the original target mask that needs to be reconstructed. In Figs. \ref{error}(b) and \ref{error}(c), the reconstructed images above a threshold level of -16 dB are represented for the CRA and MMA-based CRA configurations, respectively. In Figs. \ref{error}(d) and \ref{error}(e), the wrong reconstructed areas (red color) and correct reconstructed areas (light blue color) are represented for the CRA and MMA-based CRA configurations. The accuracy of the imaging is calculated as the number of correct reconstruction pixels divided by the total number of pixels in the imaging domain. The wrong reconstructions could be either a point in the target domain that is detected with a reflectivity level smaller than -16 dB or a point outside the target domain that has been detected with a reflectivity level higher than -16 dB. The accuracy for the CRA configuration and MMA-based CRA configurations are calculated to be 88.74\% and 92.46\%, respectively, which shows the effectiveness of the established MMAs in the imaging system. 

Finally, this improvement will be even more evident for a target that is less sparse than that used in Fig. \ref{HM_NESTA} under a Total Variation Norm metric.  Specifically, as shown in Fig. \ref{Human_Model_SVD_Capacity}, for a SNR=50 dB, the MMA-based CRA has about 30\% more singular values above the noise level than the simple CRA; so considering a typical value on the minimum number of measurements ``$m$'' needed to reconstruct an ``$s$''-sparse signal to be $m=4s$, then the MMA-based CRA should be able to reconstruct signals of sparsity $s_{CRA/MMA}=1000$, while the CRA will only be able to reconstruct signals with sparsity $s_{CRA}=750$.

\vspace{-0.2cm}

\section{Conclusion}

A new radar system for high sensing capacity imaging applications, using a metamaterial-based CRA is presented. By adopting the frequency dispersive MMAs coated on the surface of the reflector antenna, the new imaging system is able to increase the sensing capacity and create highly uncorrelated spatial and spectral patterns in the near-field of the system. Polarization-independent meander-line metamaterials are utilized as the unit-cells for designing 16 different configurations of a 4-bit MMA. Using a norm-1 regularized iterative algorithm based on ADMM, the performance of the system is evaluated in an active mm-wave imaging application. The system is characterized in terms of its radiation pattern and beam focusing capabilities. Promising results are achieved for the sensing capacity and image reconstruction quality of the 4-bit MMA-based reflector antenna, compared with that of the TRA. As a second example, an array of CRAs equipped by 8-bit MMAs was utilized to image an extended human-size region. One of the main features of the proposed CRA-based system is that it can be easily reconfigured to operate in the far-field, thus enabling standoff detection of security threats.

Although this paper only considered the metamaterials for tailoring the surface of the reflector, the use of all-dielectric materials or other MMA configurations on the surface of the reflector opens up several exciting areas of further research. In particular, utilizing an MMA with a higher quality factor allows one to fit more resonances in the frequency range of the radar. This is expected to result in a higher dispersion in the frequency response of the MMAs, which ultimately increases the sensing capacity and the imaging capabilities of the system even further.

\section*{Acknowledgment}
This work is funded by the U.S. Department of Homeland Security, Award No. “2013-ST-061-ED0001”.

\bibliography{CRA_all,ADMM}

\begin{thebibliography}{10}
\providecommand{\url}[1]{#1}
\csname url@samestyle\endcsname
\providecommand{\newblock}{\relax}
\providecommand{\bibinfo}[2]{#2}
\providecommand{\BIBentrySTDinterwordspacing}{\spaceskip=0pt\relax}
\providecommand{\BIBentryALTinterwordstretchfactor}{4}
\providecommand{\BIBentryALTinterwordspacing}{\spaceskip=\fontdimen2\font plus
\BIBentryALTinterwordstretchfactor\fontdimen3\font minus
  \fontdimen4\font\relax}
\providecommand{\BIBforeignlanguage}[2]{{%
\expandafter\ifx\csname l@#1\endcsname\relax
\typeout{** WARNING: IEEEtran.bst: No hyphenation pattern has been}%
\typeout{** loaded for the language `#1'. Using the pattern for}%
\typeout{** the default language instead.}%
\else
\language=\csname l@#1\endcsname
\fi
#2}}
\providecommand{\BIBdecl}{\relax}
\BIBdecl

\bibitem{Sheen2001}
D.~Sheen, D.~McMakin, and T.~Hall, ``Three-dimensional millimeter-wave imaging
  for concealed weapon detection,'' \emph{Microwave Theory and Techniques, IEEE
  Transactions on}, vol.~49, no.~9, pp. 1581--1592, 2001.

\bibitem{Martinez-Lorenzo2012}
J.~A. Martinez-Lorenzo, F.~Quivira, and C.~M. Rappaport, ``Sar imaging of
  suicide bombers wearing concealed explosive threats,'' \emph{Progress In
  Electromagnetics Research}, vol. 125, pp. 255--272, 2012.

\bibitem{ahmed2011novel}
S.~S. Ahmed, A.~Schiessl, and L.-P. Schmidt, ``Novel fully electronic active
  real-time millimeter-wave imaging system based on a planar multistatic sparse
  array,'' in \emph{Microwave Symposium Digest (MTT), 2011 IEEE MTT-S
  International}.\hskip 1em plus 0.5em minus 0.4em\relax IEEE, 2011, pp. 1--4.

\bibitem{lyons2013reflect}
B.~N. Lyons, E.~Entchev, and M.~K. Crowley, ``Reflect-array based mm-wave
  people screening system,'' in \emph{SPIE Security+ Defence}.\hskip 1em plus
  0.5em minus 0.4em\relax International Society for Optics and Photonics, 2013,
  pp. 890\,002--890\,002.

\bibitem{Martinez-Lorenzo2012a}
J.~A. Martinez-Lorenzo, Y.~Rodriguez-Vaqueiro, C.~M. Rappaport, O.~R. Lopez,
  and A.~G. Pino, ``A compressed sensing approach for detection of explosive
  threats at standoff distances using a passive array of scatters,'' in
  \emph{Homeland Security (HST), 2012 IEEE Conference on Technologies
  for}.\hskip 1em plus 0.5em minus 0.4em\relax IEEE, 2012, pp. 134--139.

\bibitem{jam2016horizontally}
A.~Jam and K.~Sarabandi, ``A horizontally polarized beam-steerable antenna for
  sub-millimeter-wave polarimetrie imaging and collision avoidance radars,'' in
  \emph{Antennas and Propagation (APSURSI), 2016 IEEE International Symposium
  on}.\hskip 1em plus 0.5em minus 0.4em\relax IEEE, 2016, pp. 789--790.

\bibitem{kharkovsky2007microwave}
S.~Kharkovsky and R.~Zoughi, ``Microwave and millimeter wave nondestructive
  testing and evaluation-overview and recent advances,'' \emph{IEEE
  Instrumentation \& Measurement Magazine}, vol.~10, no.~2, pp. 26--38, 2007.

\bibitem{ferris1998microwave}
D.~Ferris and N.~C. Currie, ``Microwave and millimeter-wave systems for wall
  penetration,'' in \emph{Proc. sPIE}, vol. 3375, 1998, pp. 269--279.

\bibitem{dehmollaian2008refocusing}
M.~Dehmollaian and K.~Sarabandi, ``Refocusing through building walls using
  synthetic aperture radar,'' \emph{IEEE Transactions on Geoscience and Remote
  Sensing}, vol.~46, no.~6, pp. 1589--1599, 2008.

\bibitem{dehmollaian2009through}
M.~Dehmollaian, M.~Thiel, and K.~Sarabandi, ``Through-the-wall imaging using
  differential sar,'' \emph{IEEE Transactions on Geoscience and Remote
  Sensing}, vol.~47, no.~5, pp. 1289--1296, 2009.

\bibitem{sleasman2016design}
T.~Sleasman, M.~Boyarsk, M.~F. Imani, J.~N. Gollub, and D.~R. Smith, ``Design
  considerations for a dynamic metamaterial aperture for computational imaging
  at microwave frequencies,'' \emph{JOSA B}, vol.~33, no.~6, pp. 1098--1111,
  2016.

\bibitem{ahmed2009multistatic}
S.~S. Ahmed, A.~Schiessl, and L.-P. Schmidt, ``Multistatic mm-wave imaging with
  planar 2d-arrays,'' in \emph{2009 German Microwave Conference}.\hskip 1em
  plus 0.5em minus 0.4em\relax IEEE, 2009, pp. 1--4.

\bibitem{Martinez-Lorenzo2015}
J.~Martinez-Lorenzo, J.~Heredia~Juesas, and W.~Blackwell, ``A
  single-transceiver compressive reflector antenna for high-sensing-capacity
  imaging,'' \emph{Antennas and Wireless Propagation Letters, IEEE}, vol.~PP,
  no.~99, pp. 1--1, 2015.

\bibitem{MartinezLorenzo2015}
J.~Martinez~Lorenzo, J.~Juesas, and W.~Blackwell, ``Single-transceiver
  compressive antenna for high-capacity sensing and imaging applications,'' in
  \emph{Antennas and Propagation (EuCAP), 2015 9th European Conference on},
  April 2015, pp. 1--3.

\bibitem{molaei2018interferometric}
A.~Molaei, J.~H. Juesas, W.~Blackwell, and J.~A.~M. Lorenzo, ``Interferometric
  sounding using a metamaterial-based compressive reflector antenna,''
  \emph{IEEE Transactions on Antennas and Propagation}, 2018.

\bibitem{juesas2015consensus}
J.~H. Juesas, G.~Allan, A.~Molaei, L.~Tirado, W.~Blackwell, and J.~A.~M.
  Lorenzo, ``Consensus-based imaging using admm for a compressive reflector
  antenna,'' in \emph{2015 IEEE International Symposium on Antennas and
  Propagation \& USNC/URSI National Radio Science Meeting}.\hskip 1em plus
  0.5em minus 0.4em\relax IEEE, 2015, pp. 1304--1305.

\bibitem{molaei2016interferometric}
A.~Molaei, G.~Allan, J.~Heredia, W.~Blackwell, and J.~Martinez-Lorenzo,
  ``Interferometric sounding using a compressive reflector antenna,'' in
  \emph{2016 10th European Conference on Antennas and Propagation
  (EuCAP)}.\hskip 1em plus 0.5em minus 0.4em\relax IEEE, 2016, pp. 1--4.

\bibitem{molaei2016active}
A.~Molaei, J.~H. Juesas, G.~Allan, and J.~Martinez-Lorenzo, ``Active imaging
  using a metamaterial-based compressive reflector antenna,'' in \emph{Antennas
  and Propagation (APSURSI), 2016 IEEE International Symposium on}.\hskip 1em
  plus 0.5em minus 0.4em\relax IEEE, 2016, pp. 1933--1934.

\bibitem{molaei20172}
A.~Molaei, J.~Heredia-Juesas, and J.~Martinez-Lorenzo, ``A 2-bit and 3-bit
  metamaterial absorber-based compressive reflector antenna for high sensing
  capacity imaging,'' in \emph{Technologies for Homeland Security (HST), 2017
  IEEE International Symposium on}.\hskip 1em plus 0.5em minus 0.4em\relax
  IEEE, 2017, pp. 1--6.

\bibitem{molaei2017high}
A.~Molaei, G.~Ghazi, J.~Heredia-Juesas, H.~Gomez-Sousa, and
  J.~Martinez-Lorenzo, ``High capacity imaging using an array of compressive
  reflector antennas,'' in \emph{Antennas and Propagation (EUCAP), 2017 11th
  European Conference on}.\hskip 1em plus 0.5em minus 0.4em\relax IEEE, 2017,
  pp. 1731--1734.

\bibitem{heredia2017norm}
J.~Heredia-Juesas, A.~Molaei, L.~Tirado, W.~Blackwell, and J.~{\'A}.
  Mart{\'\i}nez-Lorenzo, ``Norm-1 regularized consensus-based admm for imaging
  with a compressive antenna,'' \emph{IEEE Antennas and Wireless Propagation
  Letters}, vol.~16, pp. 2362--2365, 2017.

\bibitem{molaei2017compressive}
A.~Molaei, J.~H. Juesas, and J.~A.~M. Lorenzo, ``Compressive reflector antenna
  phased array,'' in \emph{Antenna Arrays and Beam-formation}.\hskip 1em plus
  0.5em minus 0.4em\relax InTech, 2017.

\bibitem{shannon1948mathematical}
C.~Shannon, ``A mathematical theory of communication, bell system technical
  journal 27: 379-423 and 623--656,'' \emph{Mathematical Reviews (MathSciNet):
  MR10, 133e}, 1948.

\bibitem{watts2015metamaterials}
C.~Watts, ``Metamaterials and their applications towards novel imaging
  technologies,'' Ph.D. dissertation, PhD Thesis, 2015.

\bibitem{Watts2012}
C.~M. Watts, X.~Liu, and W.~J. Padilla, ``Metamaterial electromagnetic wave
  absorbers,'' \emph{Advanced materials}, vol.~24, no.~23, pp. OP98--OP120,
  2012.

\bibitem{Ugawa1999}
A.~Ugawa, A.~G. Rinzler, and D.~Tanner, ``Far-infrared gaps in single-wall
  carbon nanotubes,'' \emph{Physical Review B}, vol.~60, no.~16, p. R11305,
  1999.

\bibitem{encinar2006dual}
J.~A. Encinar, L.~S. Datashvili, J.~A. Zornoza, M.~Arrebola,
  M.~Sierra-Casta{\~n}er, J.~L. Besada-Sanmartin, H.~Baier, and H.~Legay,
  ``Dual-polarization dual-coverage reflectarray for space applications,''
  \emph{IEEE Transactions on Antennas and Propagation}, vol.~54, no.~10, pp.
  2827--2837, 2006.

\bibitem{encinar2010recent}
J.~A. Encinar, ``Recent advances in reflectarray antennas,'' in \emph{Antennas
  and Propagation (EuCAP), 2010 Proceedings of the Fourth European Conference
  on}.\hskip 1em plus 0.5em minus 0.4em\relax IEEE, 2010, pp. 1--6.

\bibitem{lockwood1994optimizing}
G.~Lockwood and F.~S. Foster, ``Optimizing sparse two-dimensional transducer
  arrays using an effective aperture approach,'' in \emph{Ultrasonics
  Symposium, 1994. Proceedings., 1994 IEEE}, vol.~3.\hskip 1em plus 0.5em minus
  0.4em\relax IEEE, 1994, pp. 1497--1501.

\bibitem{ahmed2009near}
S.~S. Ahmed, A.~Schiess, and L.-P. Schmidt, ``Near field mm-wave imaging with
  multistatic sparse 2d-arrays,'' in \emph{2009 European Radar Conference
  (EuRAD)}, 2009.

\bibitem{byron1993radar}
E.~Byron, ``Radar principles, technology, applications,'' \emph{Englewood
  Cliffs, New Jersey: PTR Prentice hall}, pp. 618--627, 1993.

\bibitem{Meana2010}
J.~Meana, J.~{\'A}. Mart{\'\i}nez-Lorenzo, F.~Las-Heras, and C.~Rappaport,
  ``Wave scattering by dielectric and lossy materials using the modified
  equivalent current approximation (meca),'' \emph{Antennas and Propagation,
  IEEE Transactions on}, vol.~58, no.~11, pp. 3757--3761, 2010.

\bibitem{HerediaJuesas2015}
J.~Heredia~Juesas, G.~Allan, A.~Molaei, L.~Tirado, W.~Blackwell, and J.~A.
  Martinez~Lorenzo, ``Consensus-based imaging using admm for a compressive
  reflector antenna,'' in \emph{Antennas and Propagation Symposium}, July 2015.

\bibitem{degroot1974reaching}
M.~H. DeGroot, ``Reaching a consensus,'' \emph{Journal of the American
  Statistical Association}, vol.~69, no. 345, pp. 118--121, 1974.

\bibitem{erseghe2011fast}
T.~Erseghe, D.~Zennaro, E.~Dall'Anese, and L.~Vangelista, ``Fast consensus by
  the alternating direction multipliers method,'' \emph{Signal Processing, IEEE
  Transactions on}, vol.~59, no.~11, pp. 5523--5537, November 2011.

\bibitem{mota2012distributed}
J.~F. Mota, J.~Xavier, P.~M. Aguiar, and M.~Puschel, ``Distributed basis
  pursuit,'' \emph{Signal Processing, IEEE Transactions on}, vol.~60, no.~4,
  pp. 1942--1956, April 2012.

\bibitem{mota2013d}
J.~F. Mota, J.~M. Xavier, P.~M. Aguiar, and M.~Puschel, ``D-admm: A
  communication-efficient distributed algorithm for separable optimization,''
  \emph{Signal Processing, IEEE Transactions on}, vol.~61, no.~10, pp.
  2718--2723, May 2013.

\bibitem{forero2010consensus}
P.~A. Forero, A.~Cano, and G.~B. Giannakis, ``Consensus-based distributed
  support vector machines,'' \emph{The Journal of Machine Learning Research},
  vol.~11, pp. 1663--1707, 2010.

\bibitem{bredies2008linear}
K.~Bredies and D.~A. Lorenz, ``Linear convergence of iterative
  soft-thresholding,'' \emph{Journal of Fourier Analysis and Applications},
  vol.~14, no. 5-6, pp. 813--837, October 2008.

\bibitem{woodbury1950inverting}
M.~A. Woodbury, ``Inverting modified matrices,'' \emph{Memorandum report},
  vol.~42, p. 106, 1950.

\bibitem{zhou2011analysis}
M.~Zhou, S.~B. S{\o}rensen, E.~J{\o}rgensen, P.~Meincke, O.~S. Kim, and
  O.~Breinbjerg, ``Analysis of printed reflectarrays using extended local
  periodicity,'' in \emph{Antennas and Propagation (EUCAP), Proceedings of the
  5th European Conference on}.\hskip 1em plus 0.5em minus 0.4em\relax IEEE,
  2011, pp. 1408--1412.

\bibitem{Becker2011}
S.~Becker, J.~Bobin, and E.~J. Candes, ``Nesta: a fast and accurate first-order
  method for sparse recovery,'' \emph{SIAM Journal on Imaging Sciences},
  vol.~4, no.~1, pp. 1--39, 2011.

\end{thebibliography}
\bibliographystyle{IEEEtran}

\end{document}